\documentclass[a4paper, 11pt]{article}

\providecommand{\keywords}[1]
{
  \small	
  \textbf{\textit{Keywords: }} #1
}

\usepackage{amsfonts}
\usepackage{amsmath}
\usepackage{amssymb}
\usepackage{amsthm}
\usepackage{graphicx}
\usepackage{caption}
\usepackage{subcaption} 
\usepackage{float} 
\usepackage[margin=1in]{geometry} 
\usepackage[utf8]{inputenc} 
\graphicspath{ {figures/} }
\usepackage{array}
\usepackage[usenames, dvipsnames]{color} 
\usepackage{empheq} 

\usepackage{hyperref} 
\usepackage{mathtools}

\usepackage{tikz-cd} 
\usepackage{soul} 

\title{Information Shift Dynamics Described by Tsallis $q=3$ Entropy on a Compact Phase Space}

\author{
Jin Yan$^1$ and Christian Beck$^{2, 3}$ \\
\small $^1$Max Planck Institute for the Physics of Complex Systems, Dresden, Germany \\
\small \textit{jinyan@pks.mpg.de} \\
\small $^2$School of Mathematical Sciences, Queen Mary University of London, UK \\
\small $^3$The Alan Turing Institute, London, UK \\
\small \textit{c.beck@qmul.ac.uk}
}

\date{}

\begin{document}
\maketitle

\begin{abstract} 
Recent mathematical investigations have shown that under very general conditions exponential mixing implies the Bernoulli property. As a concrete example
of a statistical mechanics which is exponentially mixing we consider a Bernoulli shift dynamics by Chebyshev maps of arbitrary order $N\geq 2$, which maximizes Tsallis $q=3$ entropy rather than the ordinary $q=1$ Boltzmann-Gibbs entropy. Such an information shift dynamics may be relevant in a pre-universe before ordinary space-time is created. We
discuss symmetry properties of the coupled Chebyshev systems, which are different for even and odd $N$.
We show that the value of the fine structure constant $\alpha_{el}=1/137$ is distinguished as a coupling constant in this context, leading to uncorrelated
behaviour in the spatial direction of the corresponding coupled map lattice for $N=3$.
\end{abstract}


\section{Introduction}

The foundations of statistical mechanics are a subject of continuing theoretical interest. It is far from obvious that a given deterministic
dynamics can ultimately  be described by a statistical mechanics formalism. The introduction of generalized entropies (such as, for example,
the non-additive Tsallis entropies \cite{nonex, tsallis-book,jizba-korbel}) leads to a further extension of the formalism relevant for systems with long-range interactions, or with a fractal or compactified phase space structure, or with non-equilibrium
steady states exhibiting fluctuations of temperature or of effective diffusion constants \cite{super, metzler}. Generalized entropies have been shown to have applications for a variety of complex systems, for example, in high energy physics \cite{curado, beck-2000, biro, deppman, wilk} or for turbulent systems \cite{turbu, beck-review, yoon}. In this paper we go back to the basics and explore the properties of a particular statistical mechanics, namely that of an information
shift dynamics described by Tsallis entropies with the entropic index $q=3$ on a compact phase space.

Our work is inspired by recent mathematical work \cite{dd21} that shows that the exponential mixing property automatically implies the
Bernoulli property, under very general circumstances. This theorem is of utmost interest for the foundations of statistical mechanics.
Namely, by definition the systems we are interested in when dealing with statistical mechanics relax to an equilibrium quite quickly (under normal circumstances). That is to say, we
have the exponential mixing property. But then this implies that somewhere (on a suitable subset of the phase space) there must exist a Bernoulli shift
dynamics in suitable coordinates.

We will work out one of the simplest example systems that is consistent with a generalized statistical mechanics formalism, and at the same time
is exponentially mixing and ultimately conjugated to a Bernoulli shift. These are discrete-time dynamical systems on the interval $[-1,1]$ as generated by
$N$-th order Chebyshev maps $T_N$ \cite{jycb20, hilgers,groote,dettmann}. Chebyshev maps are exponentially mixing and conjugated to a Bernoulli shift of $N$ symbols. We will review and
investigate their properties in detail in the following sections. Needless to say that Chebyshev maps do not live in ordinary physical space
but just on a compactified space, the interval $[-1,1]$. The simplest examples are given by the $N=2$ and $N=3$ cases, i.e., $T_2(x)=2x^2-1$ and $T_3(x)=4x^3-3x$. Despite their simplicity, a statistical mechanics formalism can be constructed as an effective description (see also \cite{book, escort}). In contrast
to ordinary statistical mechanics (described by states where the $q=1$ Boltzmann-Gibbs-Shannon entropy has a maximum subject to constraints),
in our case the relevant entropy is the $q=3$ Tsallis entropy $S_q$. This leads to interesting properties. Our physical interpretation
is that the above information shift dynamics may be relevant in a pre-universe, i.e. in an extremely early stage of the universe
where ordinary space-time has not yet formed. The dynamics evolves in a fictitious time coordinate (different from physical time) which
is relevant for stochastic quantization \cite{stoch1, stoch2} (this idea has been worked out in more detail in \cite{book, physica-d,prd}).

 While we will work out the properties of Chebyshev maps in much detail in the following sections,
 what we mention right now as a prerequisite is that the invariant density $p(x)$ of Chebyshev maps $T_N$ of arbitrary order $N \geq 2$ is given by
\begin{equation}
p(x) = \frac{1}{\pi \sqrt{1-x^2}},\;\; x \in [-1,1]. \label{inva}
\end{equation}
 This density describes the probability density of iterates under long-term iteration.
 The maps $T_N$ are ergodic and mixing. In suitable coordinates iteration of the map $T_N$ corresponds to a Bernoulli shift dynamics of $N$ symbols. In the following we give a generalized statistical mechanics interpretation for the above invariant density, identifying Chebyshev maps as one of the simplest systems possible for which a generalized statistical mechanics can be defined. The interesting aspect of this low-dimensional simplicity is that $q=3$ is relevant,
 rather than $q=1$ as for ordinary statistical mechanics.
 
 This paper is organized as follows. In section \ref{sect-cano} we derive the generalized canonical distributions
 obtained from the $q$-entropies $S_q$, and discuss the special distinguished features obtained for $q=3$ (or $q=-1$ if
 so-called escort distributions \cite{escort,mendes} are used). In section \ref{sect-cheby} we discuss the exponential mixing property, which is fixed by the
 2nd largest eigenvalue of the Perron-Frobenius operator. In fact, we will present formulas for {\em all} eigenvalues and eigenfunctions,
 thus completely describing the exponential mixing behaviour. In section \ref{sect-cml} we couple two maps, thus gradually expanding the phase space, and investigate how the coupling structure
 induces certain symmetries in the attractor which are different for odd $N$ and even $N$. The degeneracy of the canonical distribution
 (of the invariant density) is removed by the coupling, and all attractors become $N$-dependent. Finally, in section \ref{sect-snnc} we consider a large number of coupled maps on a one-dimensional lattice space. This is the realm of the so-called 'chaotic strings' which have previously been shown
 to have relevant applications in quantum field theory \cite{groote,book,physica-d}. We confirm, by numerical simulations, that the $N=3$ string distinguishes a value
 of the coupling constant given numerically by 1/137, which numerically coincides with the low-energy limit of the fine structure constant fixing the strength of electric interaction.
 A physical interpretation for that will be given, in the sense that we assume that the chaotic shift dynamics, described by $q=3$
 Tsallis entropies in a statistical mechanics setting, is a fundamental information shift dynamics that helps to fix standard model parameters
 in a pre-universe, before ordinary space-time is created.

\section{Generalized canonical distributions from maximizing $q$-entropy subject to constraints}
\label{sect-cano}

The relevant
probability density (\ref{inva}) mentioned above can be regarded as a generalized canonical
distribution in non-extensive statistical mechanics \cite{nonex}.
As it is well known, one defines for a
dimensionless continuous random variable $X$ with probability density
$p(x)$ the
Tsallis entropies as
\begin{equation}
S_q=\frac{1}{q-1}(1-\int p(x)^qdx).\label{sq}
\end{equation}
Here $q\in (-\infty , \infty)$ is the entropic index. The Tsallis entropies contain the
Boltzmann-Gibbs-Shannon entropy $S_1=-\int p(x) \log p(x)dx$ as a special case for $q\to 1$,
as can be easily checked by writing $q=1+\epsilon$ and taking the limit $\epsilon \to 0$ in the above equation. 

We now do statistical mechanics for general $q$. Typically one
has some knowledge on the system. This could be, for example, a knowledge of the mean energy $U$ of the system. Extremizing
$S_q$ subject to the constraint
\begin{equation}
\int p(x)E(x)dx =U \label{constraint}
\end{equation}
one ends up with $q$-generalized canonical distributions (see, e.g. \cite{tsallis-book, beck-review} for a review). These
are given by
\begin{equation}
p(x)\sim (1+(q-1)\beta E(x))^{-\frac{1}{q-1}}, \label{can}
\end{equation}
where $E$ is the energy associated with microstate $x$,
and $\beta =1/kT$ is the inverse temperature. Of course,
for $q\to 1$ one obtains the usual Boltzmann factor $e^{-\beta
E}$.

Alternatively, one can work with the so-called escort distributions, defined for a given parameter $q$ and a given distribution $p(x)$ as
\cite{escort}
\begin{equation*}
P(x)=\frac{p(x)^q}{\int p(x)^q dx}. 
\end{equation*}
The escort distribution sometimes helps to avoid diverging integrals, thus `renormalizing' the theory under consideration, see \cite {beck-escort} for more details.
If the energy constraint (\ref{constraint}) is implemented using the escort
distribution $P(x)$ rather than the original distribution $p(x)$,
one obtains generalized canonical distributions of the form
\begin{equation}
P(x)\sim (1+(q-1)\beta E(x))^{-\frac{q}{q-1}}. \label{Can}
\end{equation}
Again the limit $q\to 1$ yields ordinary Boltzmann factors
$e^{-\beta E}$. 

Let us now apply a $q$-generalized statistical mechanics formalism to the chaotic dynamics $x_{n+1}=T_N (x_n)$ as generated by Chebyshev maps. 
We may regard these Chebyshev maps as the simplest possibility to microscopically generate a dynamics that exactly follows a generalized canonical distribution, and at the same time lives on a compact
phase space $[-1,1]$. We
might identify $E=\frac{1}{2}mX^2$ as a formal kinetic energy
associated with the chaotic fields $X$, interpreting
$X$ as a kind of velocity and $m$ as a mass. Or, we might regard it as a harmonic oscillator potential, regarding $X$ as a kind of position. We then get by comparing
eq.~(\ref{inva}) and (\ref{can})
\begin{eqnarray*}
q&=&3\\ \beta^{-1}&=&-m
\end{eqnarray*}
Note that the mass that formally comes
out of this approach is negative. 
A formal problem of this approach is that
the $q=3$ Tsallis entropy of the distribution (\ref{inva})
as defined by the integral eq.~(\ref{sq}) does
not exist, since the integral $\int_{-1}^1 (1-x^2)^{-3/2}dx$
diverges. This problem, however, can be avoided by proceeding to the escort distribution, which has a different effective $q$ index. For escort distributions, all relevant integrals are well-defined, as we will show in the following.

If the
escort formalism \cite{escort, mendes, beck-escort} is used, then by comparing the functional form of the invariant density $\pi^{-1} (1-x^2)^{-\frac{1}{2}}$ with 
eq.~(\ref{Can}) we get
\begin{eqnarray*}
q&=& -1 \\ \beta^{-1} &=&m.
\end{eqnarray*}
We obtain the result that our shift-of-information model behaves similar to
an ideal gas in the non-extensive formalism but with entropic index
either $q=3$ or $q=-1$, rather than $q=1$ as in ordinary
statistical mechanics. The 'velocity' $v$ is given by the
dimensionless variable $v=X$, the 'kinetic energy' by the
non-relativistic formula $E=\frac{1}{2}mv^2$, and the inverse
temperature is $\beta =m^{-1}$. This non-extensive gas has the
special property that the temperature coincides with the mass of
the 'particles' considered. Alternatively, it corresponds to a harmonic oscillator potential if $X$ is 
interpreted as a fluctuating position variable
that lives on a compact interval. The idea can also be worked out for coupled systems, i.e., for
spatially coupled Chebyshev maps on a lattice, and applied to
vacuum fluctuations in quantum field theory,
see \cite{book} for more details.

We can now evaluate all interesting thermodynamic properties of
the system using the escort formalism. Regarding the invariant
density of our system as an escort distribution $P(x)=\frac{1}{\pi} (1-x^2)^{-\frac{1}{2}}$, we have, by comparing the exponent with that in eq.~(\ref{Can}), $q=-1$. The
original distribution $\hat{p}(x)$ is then given by $\hat{p}(x)\sim P(x)^{\frac{1}{q}}=P(x)^{-1}$ which gives
\begin{equation}
\hat{p}(x)=\frac{2}{\pi}\sqrt{1-x^2}, \label{pphi}
\end{equation}
where the prefactor is fixed by the normalisation condition $\int_{-1}^1 \hat{p}(x)dx = 1$. 
For the Tsallis entropy of the chaotic fields we obtain from
eq.~(\ref{sq}) and (\ref{pphi})
\begin{equation*}
S_{q}[\hat{p}]=\frac{1}{4}(\pi^2 -2)=S_{q}[P] 
\end{equation*}
It is invariant under the transformation $\hat{p}\to P$, which is a distinguished property
of the entropic index $q=\pm 1$.
For the generalized internal energy we obtain
\begin{equation*}
U_{q}[P]=\int_{-1}^1P(x)\frac{1}{2}mx^2 dx =\frac{1}{4} m
\end{equation*}
and for the generalized free energy
\begin{equation*}
F_q=U_q-TS_q=\frac{m}{4}(3-\pi^2) .
\end{equation*}

All expectations formed with the invariant measure can be regarded
as corresponding to escort expectations within the more general
type of thermodynamics that is relevant for our chaotically
evolving information-shift fields. Whereas ordinary statistical physics is described by a
statistical mechanics with $q=1$, the chaotic fields underlying our information shift
are described by $q=-1$ or $q=3$, depending
on whether the escort formalism is used or not. 
In general, one can
easily verify that the entropic indices $q=+ 1$ and $q=-1$ are
very distinguished cases: Only for these two cases the Tsallis entropy
of the escort distribution is equal to the Tsallis entropy of the
original distribution, for arbitrary distributions $p(x)$. 

An important distinction comes from the non-additivity
property of Tsallis entropies. Consider a
factorization of probabilities, i.e., write
\begin{equation*}
    p(x)=p^{I}(x) p^{II}(x)
\end{equation*}
where $I$ and $II$ are independent subsystems.
A general property of Tsallis entropy is
\begin{equation*}
    S_q^{I,II}= S_q^I +S_q^{II} -(q-1) S_q^I \cdot S_q^{II}
\end{equation*}
which is true for arbitrary factorizing probability
densities and arbitrary $q$. 

The cases $q=-1$ and $q=3$ are distinguished as they satisfy the relation
\begin{equation*}
S_q^{I,II}-S_q^I -S_q^{II}=(S_q^I \pm S_q^{II})^2-(S_q^{I})^2 -(S_q^{II})^2.
\end{equation*}
That is to say, linear entropy differences on the left-hand side are connected with squared entropy differences on the right-hand side.

We should mention here that the case $q=3$, respectively $q=-1$ is a kind of extreme case where $|q-1|$ is not small. Usually, the difference $|q-1|$ is assumed to be small in typical applications of Tsallis statistics to cosmology and quantum field theory, see e.g. \cite{luciano, luciano-blasone, nojiri} for some recent work in this direction. Note that for our information shift dynamics the Kolmogorov-Sinai entropy $h_{KS}$ is still given by the standard Lyapunov exponent $h_{KS}=\log N>0$ as we have a strongly chaotic dynamics for which everything can be calculated analytically since the invariant density is known. This means there is on average a linear growth of information in each iteration step \cite{escort}. 
All $n$-point higher-order correlation functions of the dynamics can be calculated by implementing a graph-theoretical method developed in \cite{hilgers}. The deformation from ordinary statistical mechanics is visible in the shape of the invariant density which strongly differs from a Gaussian --- it is described by a $q$-Gaussian with $q=1 \pm 2$. This anomaly is due to the fact that the statistical mechanics we consider has as a phase space a compactified space --- it just lives on the interval [-1,1]. Our physical interpretation is that this information shift dynamics is not living in the ordinary space-time of the current universe but describes a pre-universe dynamics on a compactified space. Only later, when ordinary space-time unfolds, $q$ starts getting closer to 1, because then the available phase space becomes much bigger and many more degrees of freedom will start to contribute. We may assume that there is a complex transition scenario from $q=3$ to $q=1$ on that way towards ordinary statistical mechanics.

\section{Relaxation properties of Chebyshev dynamical systems}
\label{sect-cheby}
Formally, the Chebyshev maps are a family of discrete dynamical systems defined by 
\begin{equation}
x_{n+1} = T_N(x_n), \quad n = 0, 1, 2, ..., 
\end{equation}
where 
\begin{equation}
T_N(x) = \cos (N\arccos x), \quad x \in [-1, 1], \quad N = 2, 3, 4, ..., 
\end{equation}
which can be written as polynomials, for example, 
\begin{equation}
\begin{split}
T_2(x) &= 2x^2 - 1, \\
T_3(x) &= 4x^3 - 3x, \\
T_4(x) &= 8x^4 - 8x^2 + 1, \\
&\vdots 
\end{split}
\end{equation}
They are proven to be topologically conjugated to tent-like maps via the conjugacy function $h(x) = \frac{1}{\pi}\arccos(-x)$, and as a consequence, they are conjugated to Bernoulli shifts of $N$ symbols. 
Furthermore, \cite{jycb20} showed that the (higher-order) correlation functions of Chebyshev maps are vanishing identically, except
for special sets of tuples
described by appropriate $N$-ary graphs \cite{jycb20, hilgers}, which is a distinguished feature compared to other maps conjugated to a Bernoulli shift.

The transfer (or Perron-Frobenius) operator $\mathcal{L}$ for a given map describes the time evolution of a distribution of points characterised by a density function. Consider the eigenvalue equation $\mathcal{L}\rho = \lambda \rho$: if we write an initial distribution as a linear combination of eigenfunctions of $\mathcal{L}$, then the time evolution of the distribution is just applications of the (linear) transfer operator, which is simply  multiplying the eigenfunctions in the expansion of the initial density with the corresponding eigenvalues. 
\\
For a one-dimensional discrete dynamical system $T$ the Perron-Frobenius operator is given by $\mathcal{L}\rho(y) = \sum_{x \in T^{-1}(y)} \frac{\rho(x)}{|T'(x)|}$, where $T'$ denotes the derivative of the map $T$ and $\sum_{x \in T^{-1}(y)}$ is summing over all the preimages of a point $y$. It was shown \cite{jycb20} that for Chebyshev maps $T = T_N$, we have $\mathcal{L}\rho^{(n)}(x) = \lambda^{(n)}\rho^{(n)}(x)$, $n = 0, 1, 2, ...$ with 
\begin{equation*}
\begin{split}
\lambda^{(n)} &= N^{-2n}, \\
\rho^{(n)}(x) &= \begin{cases}
\frac{1}{\pi \sqrt{1 - x^2}} B_{2n}\left( \frac{1}{2\pi}\arccos(-x) + \frac{1}{2}\right) \text{ if $N = 2, 4, ...$ is even}; \\
\begin{cases}
\rho^{(n, 1)}(x) = \frac{1}{\pi \sqrt{1 - x^2}}B_{2n}\left( \frac{1}{\pi}\arccos(-x)\right) \\
\rho^{(n, 2)}(x) = \frac{1}{\pi \sqrt{1 - x^2}}E_{2n-1}\left( \frac{1}{\pi} \arccos(-x)\right)
\end{cases} \text{ if $N = 3, 5, ...$ is odd}; 
\end{cases}
\end{split}
\end{equation*} 
where $B_n(x)$ and $E_n(x)$ are Bernoulli and Euler polynomials, respectively. 
The graphs of the first few eigenfunctions are shown below. They form a basis of the functional space under consideration.
\begin{figure}[H]
\captionsetup[subfigure]{justification=centering}
\centering 
\begin{subfigure}{0.32\linewidth}
	\includegraphics[width = \linewidth]{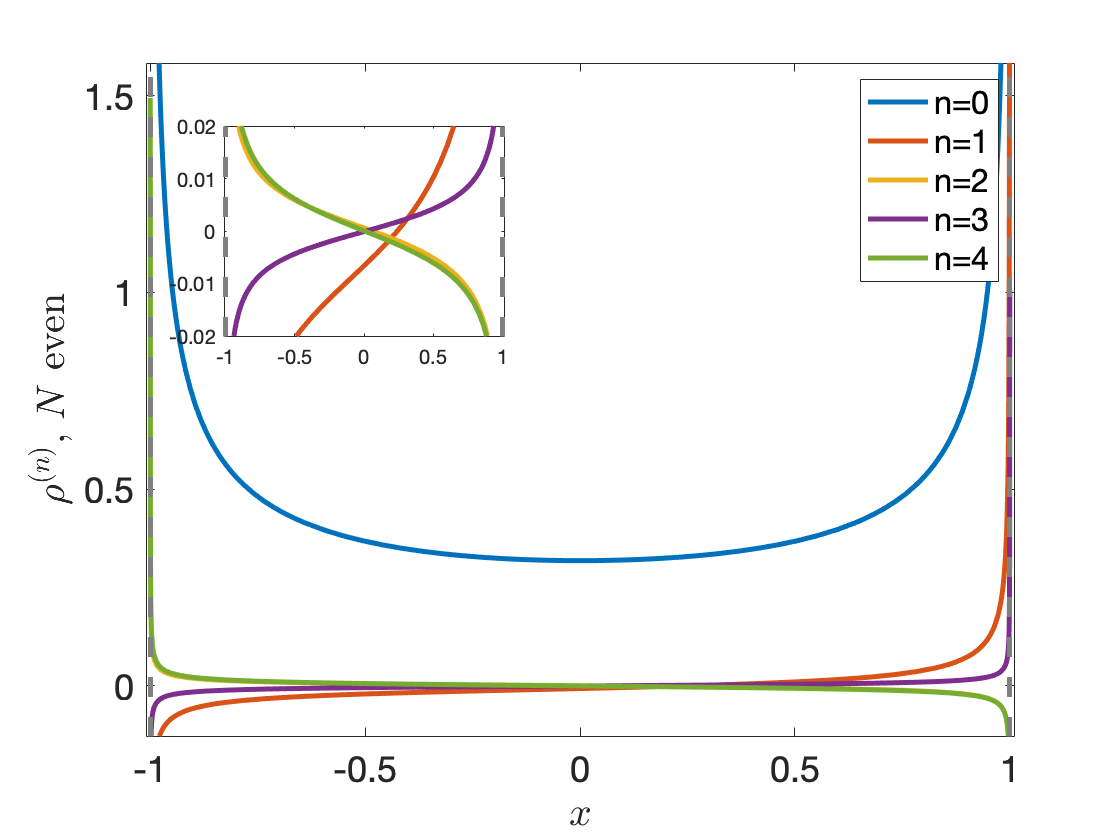}
	\caption{}\label{efun-Nodd}
\end{subfigure}
\begin{subfigure}{0.32\linewidth}
	\includegraphics[width = \linewidth]{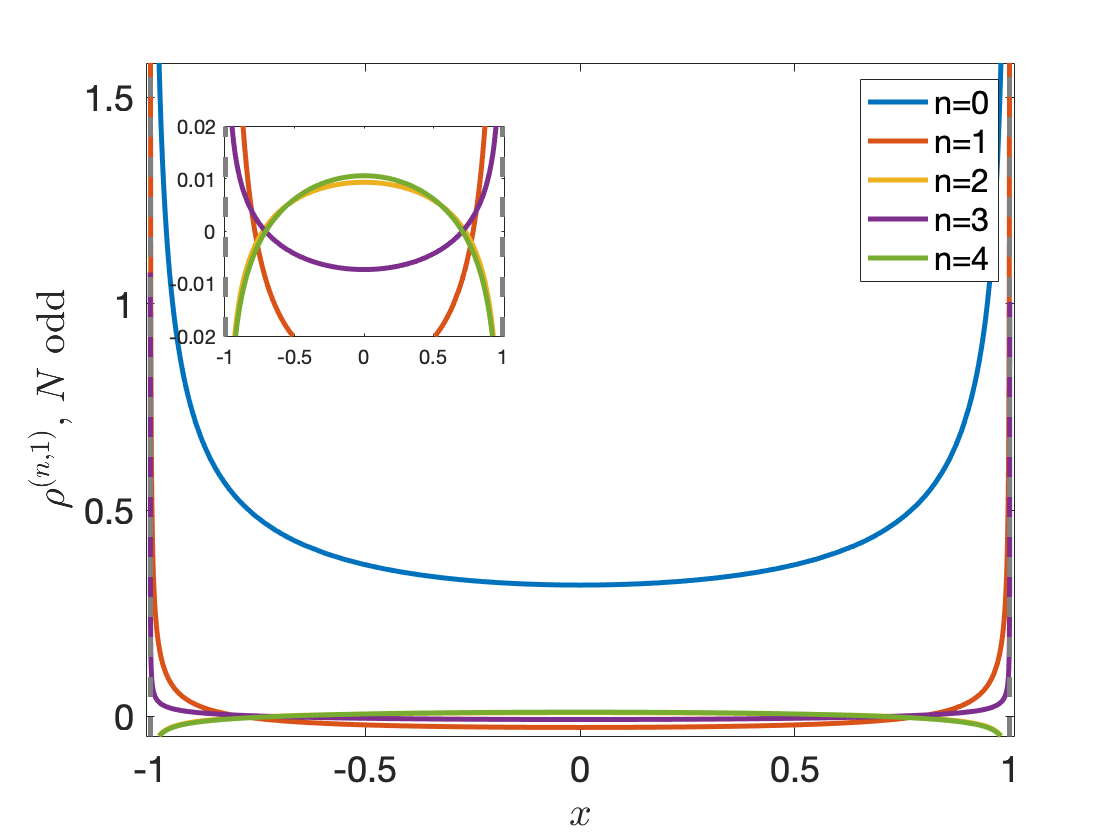} 
	\caption{}\label{efun-Neven1}
\end{subfigure}
\begin{subfigure}{0.32\linewidth}
	\includegraphics[width = \linewidth]{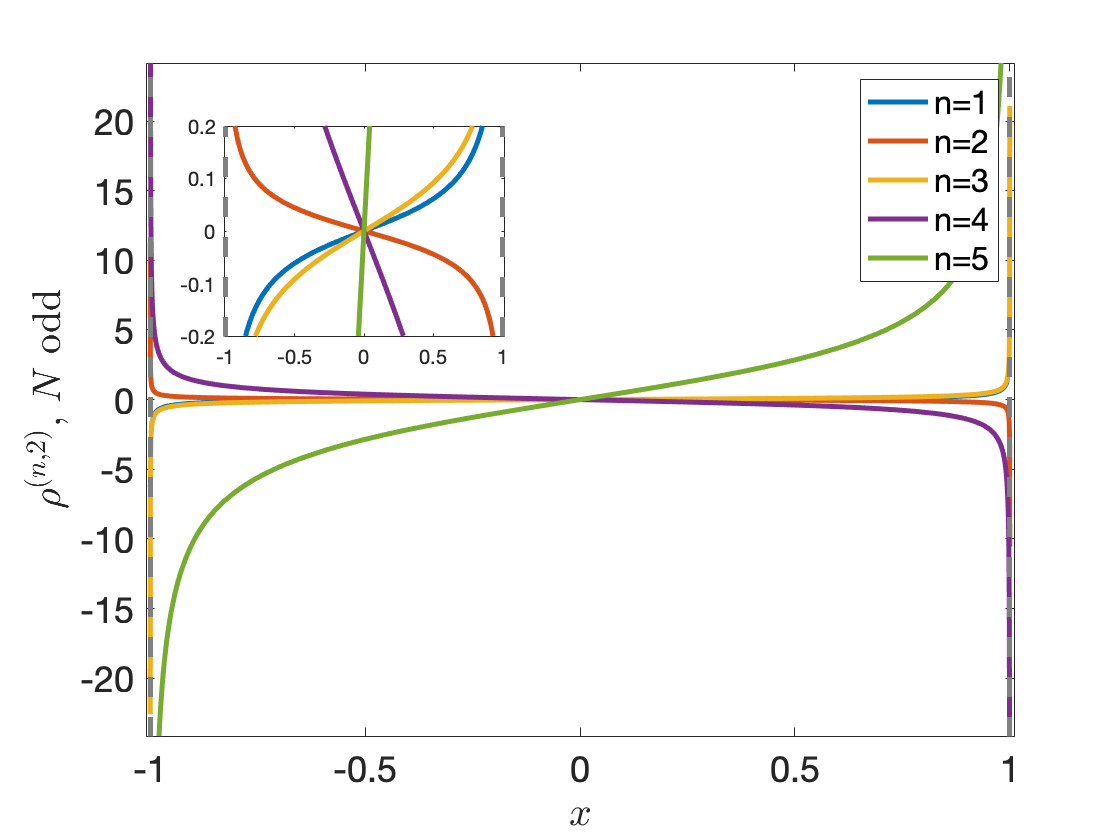} 
	\caption{}
\end{subfigure}
\caption{First five eigenfunctions of the transfer operator $\mathcal{L}$ for Chebyshev maps $T_N$ ($N \in \mathbb{N}_{\geq 2}$) with \textbf{(a)} even $N$, and \textbf{(b)}, \textbf{(c)} odd $N$. Each inset captures the graph close to the origin. \label{efuns}}
\end{figure} 

The invariant density of the Chebyshev map $T_N$ is the eigenfunction of the transfer operator $\mathcal{L}$ associated with the unit eigenvalue, with $\lambda^{(0)} = 1$ we recover eq.\eqref{inva}: $\rho^{(0)}(x) = \frac{1}{\pi \sqrt{1 - x^2}} = p(x)$. The graph can be found in blue in Fig.\ref{efun-Nodd} or Fig.\ref{efun-Neven1}. 

The remaining eigenvalues are related to relaxation times of the dynamical system and in particular, the second one $\lambda^{(1)} = N^{-2} < 1$ characterises the mixing rate in the following sense: 
\\
Consider an arbitrary initial distribution $\rho_0(x)$ in a suitable functional space, it can then be expressed as a linear combination of the eigenfunctions of $\mathcal{L}$
\begin{equation*}
\rho_0(x) = \sum_{i = 0}^{\infty} c_i \rho^{(i)}(x),
\end{equation*}
where the $c_i$ are some real coefficients. Applying the transfer operator $\mathcal{L}$ $m$ times gives 
\begin{equation*}
\rho_m(x) = \mathcal{L}^m \rho_0(x) = \sum_{i = 0}^{\infty}c_i (\lambda^{(i)})^m\rho^{(i)}(x) = c_0 p(x) + R_m, 
\end{equation*}
where $p(x)$ is the invariant density and $R_m := \sum_{i = 1}^{\infty}c_i (\lambda^{(i)})^m\rho^{(i)}(x)$ denotes the sum of remaining terms. Since $|\lambda^{(i)}| = N^{2i} < 1$ for all $i = 1, 2, ...$ and $N \in \mathbb{N}_{\geq 2}$, we have $R_m \to 0$ as $m \to \infty$. The decay of $R_m$ will be dominated by the second largest eigenvalue. In other words, $\lambda^{(1)}$ determines the rate of approaching the equilibrium state (or invariant density) $p(x)$.  

The {\it exponential mixing} property of a map means that an arbitrary density approaches the invariant density exponentially fast: $|\rho_m(x) - p(x)| \sim e^{-rm}$, where $r^{-1}$ is referred as {\it relaxation time} and it is related to the second largest eigenvalue $\lambda^{(1)}$ by $e^{-r} = |\lambda^{(1)}|$\cite{escort}. 
In our case we have $\lambda^{(1)} = N^{-2}$ and hence the relaxation time for the Chebyshev map $T_N$ is given by $r^{-1} = (2\ln N)^{-1}$.

\section{Symmetries in the attractors of coupled Chebyshev map systems of small size} 
\label{sect-cml}
We have shown in the previous section that the invariant density of a Chebyshev map $T_N$ is independent of the order $N \in \mathbb{N}_{\geq 2}$. We will now couple the maps \cite{groote,dettmann,book,physica-d,kaneko, mackey}. The independence of $N$ will be removed by the coupling and all the attractors will become $N$-dependent. 

Consider a one-dimensional lattice of size $J \geq 2$ with periodic boundary conditions, and on each lattice site we impose an identical Chebyshev dynamics. A simplest coupling scheme is the nearest-neighbour coupling and it is described as follows: 
\begin{equation}
x^{(j)}_{n+1} = (1-c)T_N(x^{(j)}_n) + \frac{c}{2}\left[ T_N(x^{(j-1)}_n) + T_N(x^{(j+1)}_n)\right],
\label{cml-eq}
\end{equation}
where the superscripts $j = 1, 2, ..., J$ and subscripts $n = 0, 1, 2, ...$ of the dynamical variable $x$ denote the spatial position of the lattice site and the number of time steps of iterations, respectively, and $c \in [0, 1)$ is the coupling strength. 

It is usually assumed that for a very weak coupling, ergodicity still holds if the uncoupled system is ergodic. It is still an open mathematical question whether the coupled Chebyshev systems are ergodic or not for a range of the coupling strength for a given order $N$. 
However, we can numerically investigate their behaviour for large time. 

\begin{figure}[H]
\captionsetup[subfigure]{justification=centering}
\centering 
\begin{subfigure}{0.24\linewidth}
	\includegraphics[width = \linewidth]{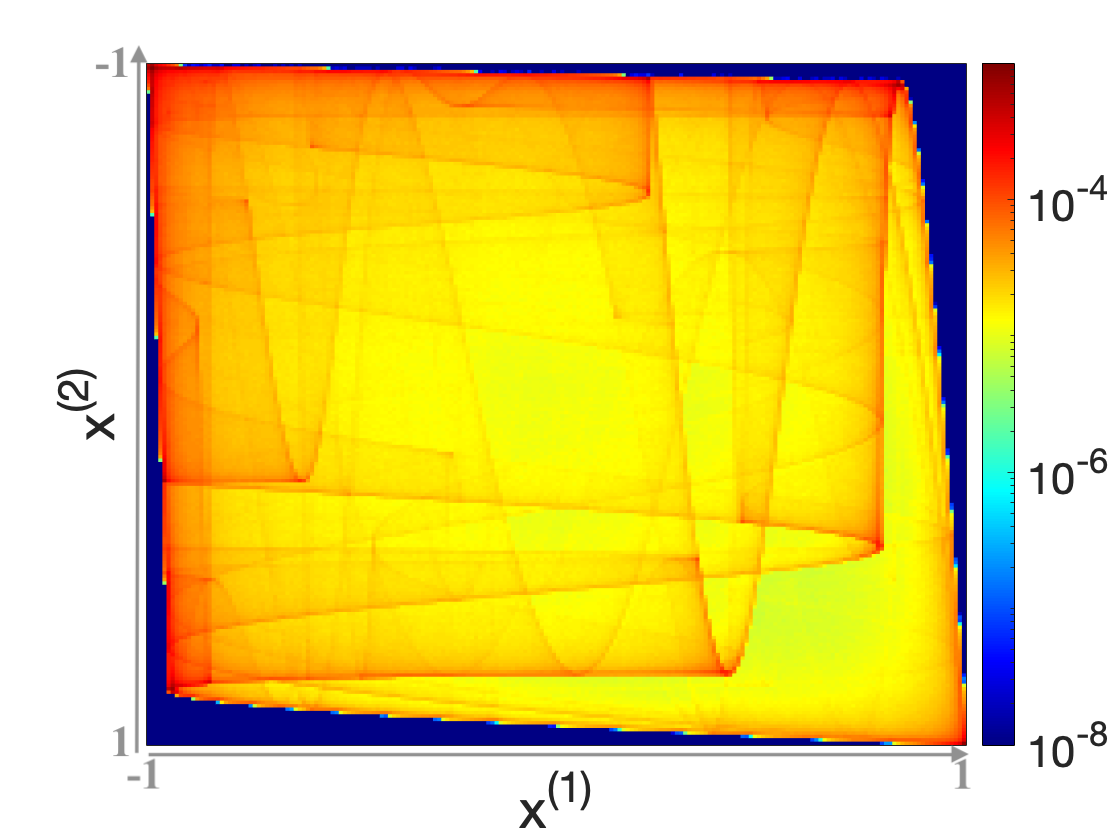}
	\caption{$T_2$: $c = 0.027$}
\end{subfigure}
\begin{subfigure}{0.24\linewidth}
	\includegraphics[width = \linewidth]{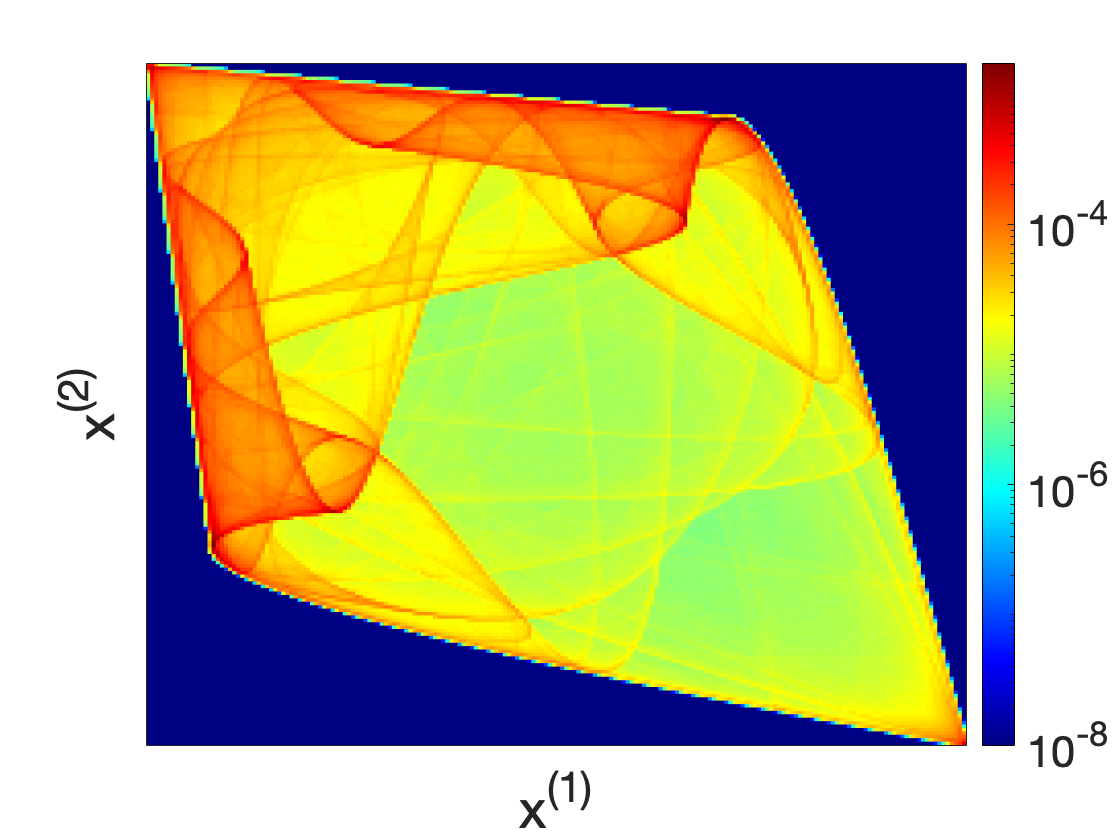} 
	\caption{$T_2$: $c = 0.1$}
\end{subfigure}
\begin{subfigure}{0.24\linewidth}
	\includegraphics[width = \linewidth]{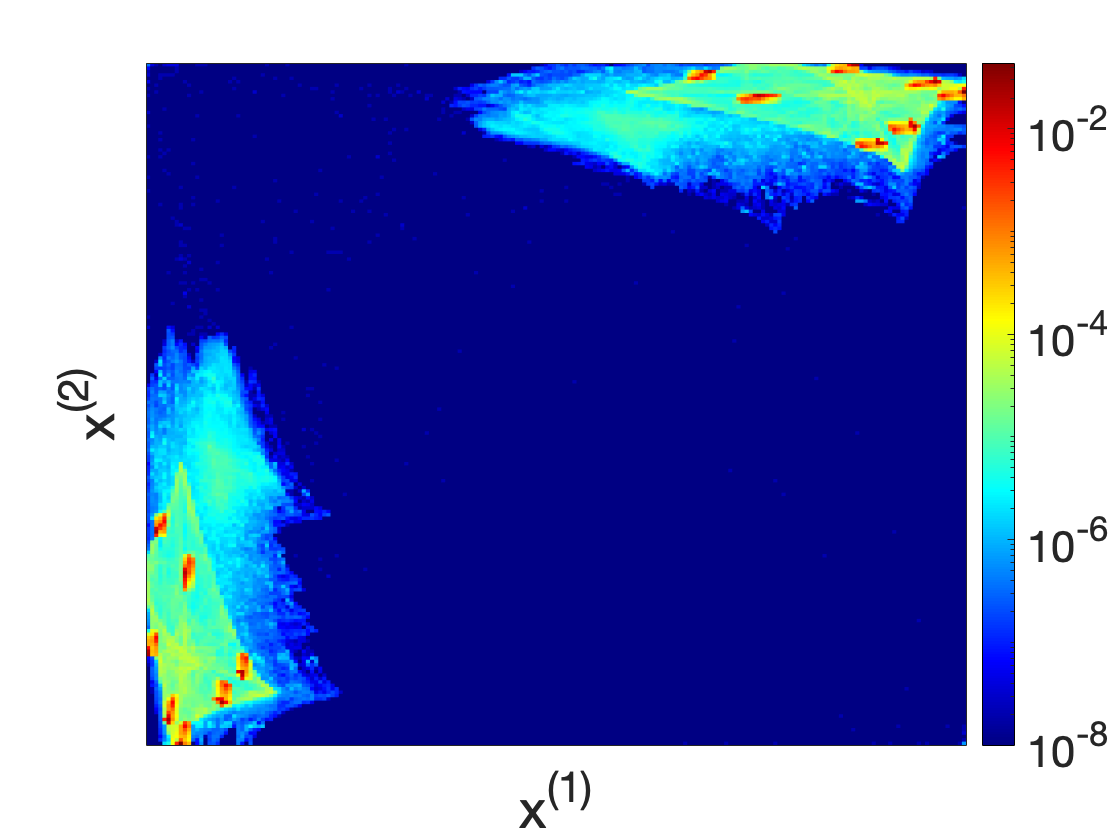} 
	\caption{$T_2$: $c = 0.12$}
\end{subfigure}
\begin{subfigure}{0.24\linewidth}
	\includegraphics[width = \linewidth]{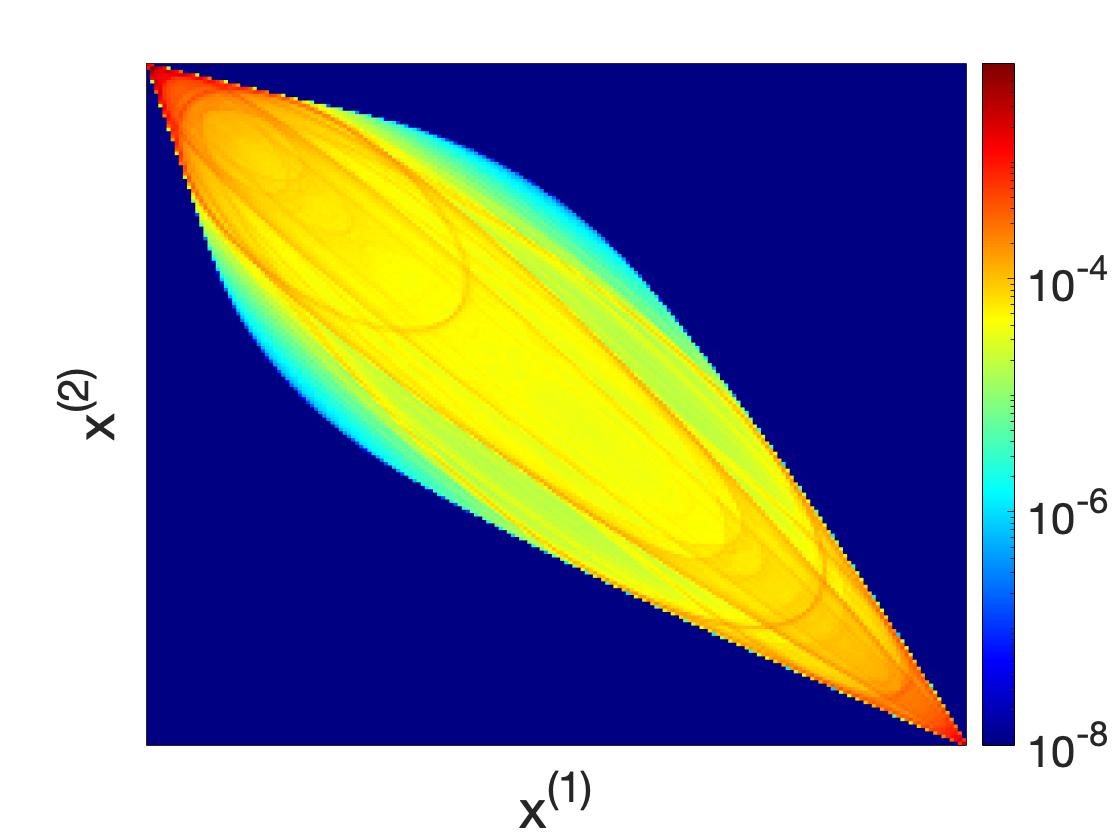} 
	\caption{$T_2$: $c = 0.22$}
\end{subfigure}
\vspace{1em}

\begin{subfigure}{0.24\linewidth}
	\includegraphics[width = \linewidth]{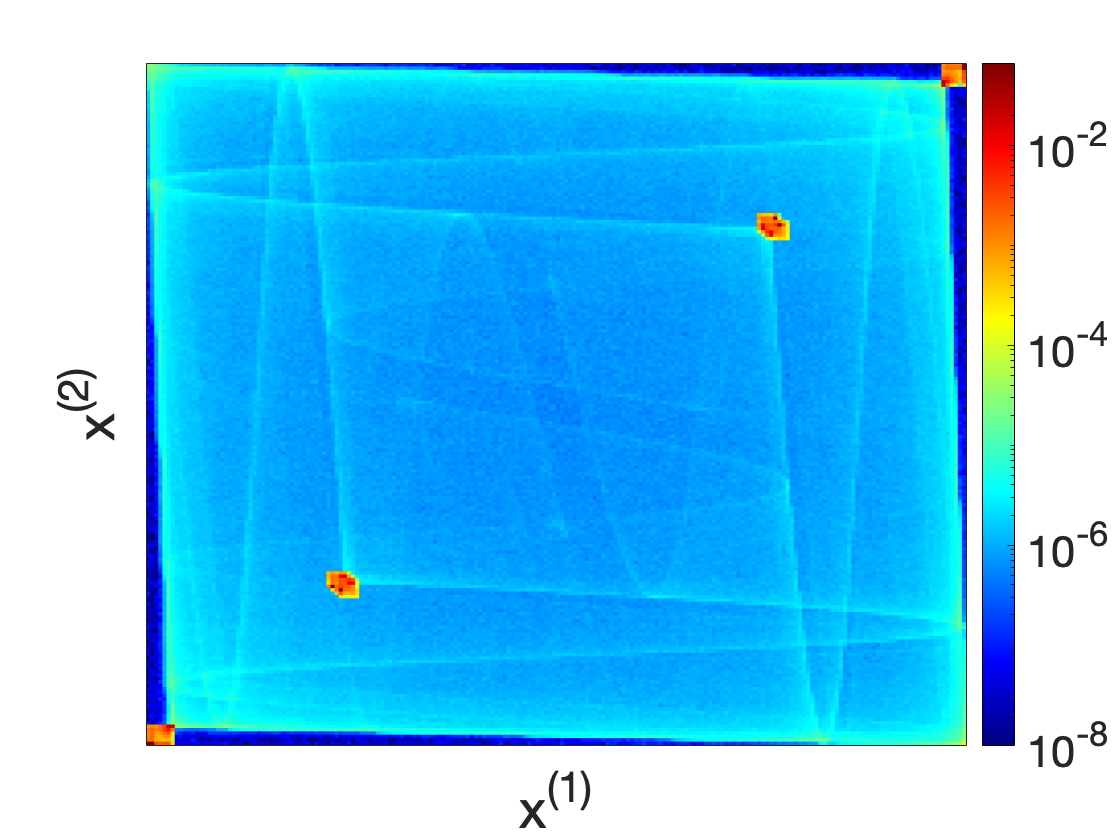}
	\caption{$T_3$: $c = 0.027$}
\end{subfigure}
\begin{subfigure}{0.24\linewidth}
	\includegraphics[width = \linewidth]{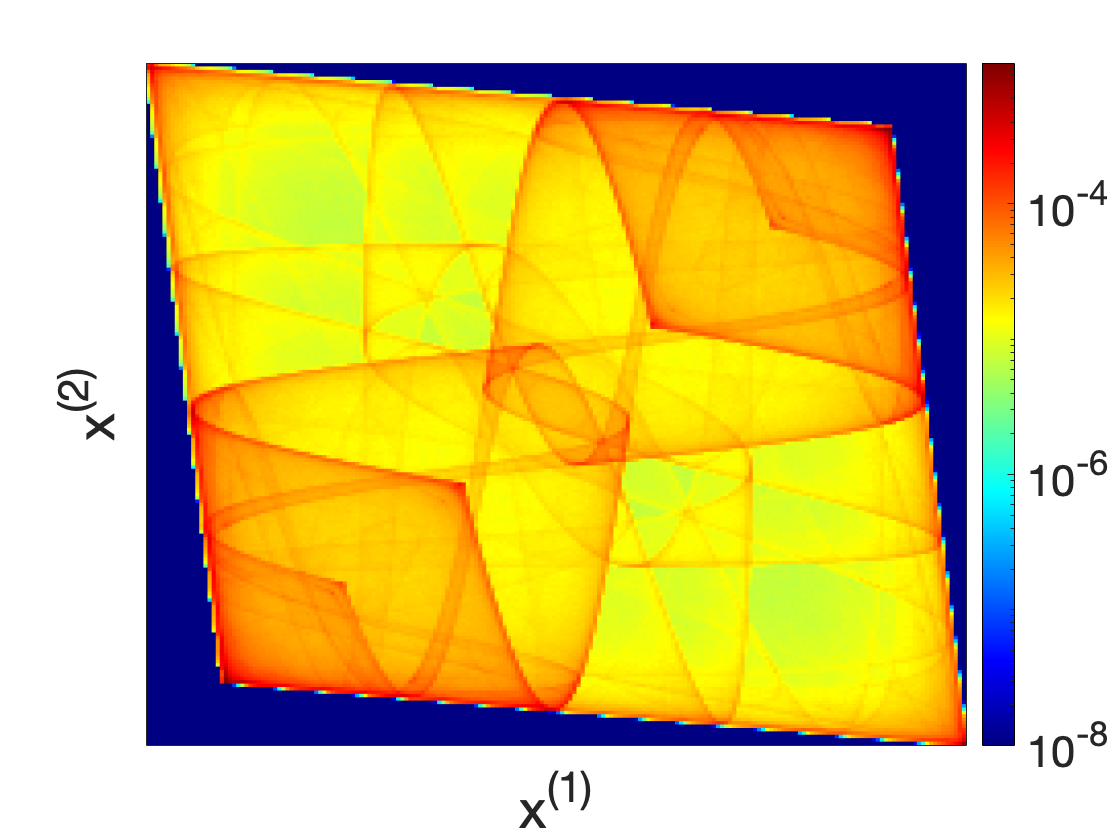} 
	\caption{$T_3$: $c = 0.093$}
\end{subfigure}
\begin{subfigure}{0.24\linewidth}
	\includegraphics[width = \linewidth]{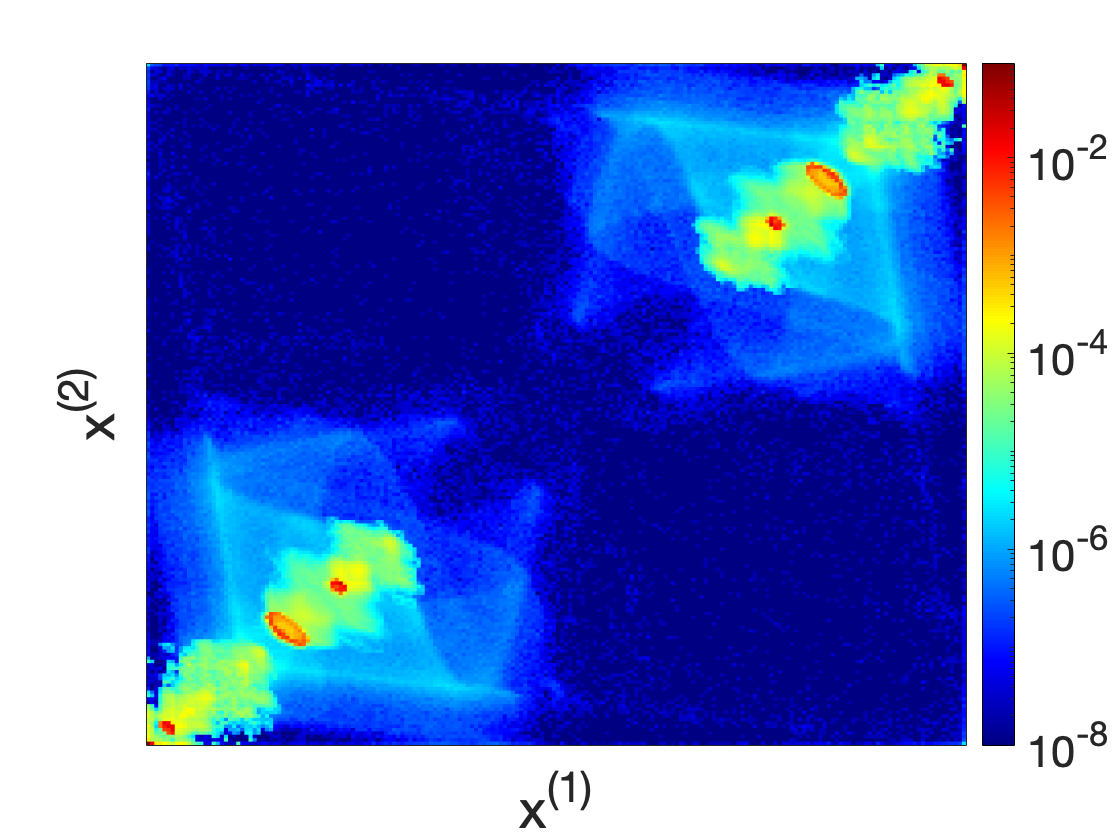} 
	\caption{$T_3$: $c = 0.12$}
\end{subfigure}
\begin{subfigure}{0.24\linewidth}
	\includegraphics[width = \linewidth]{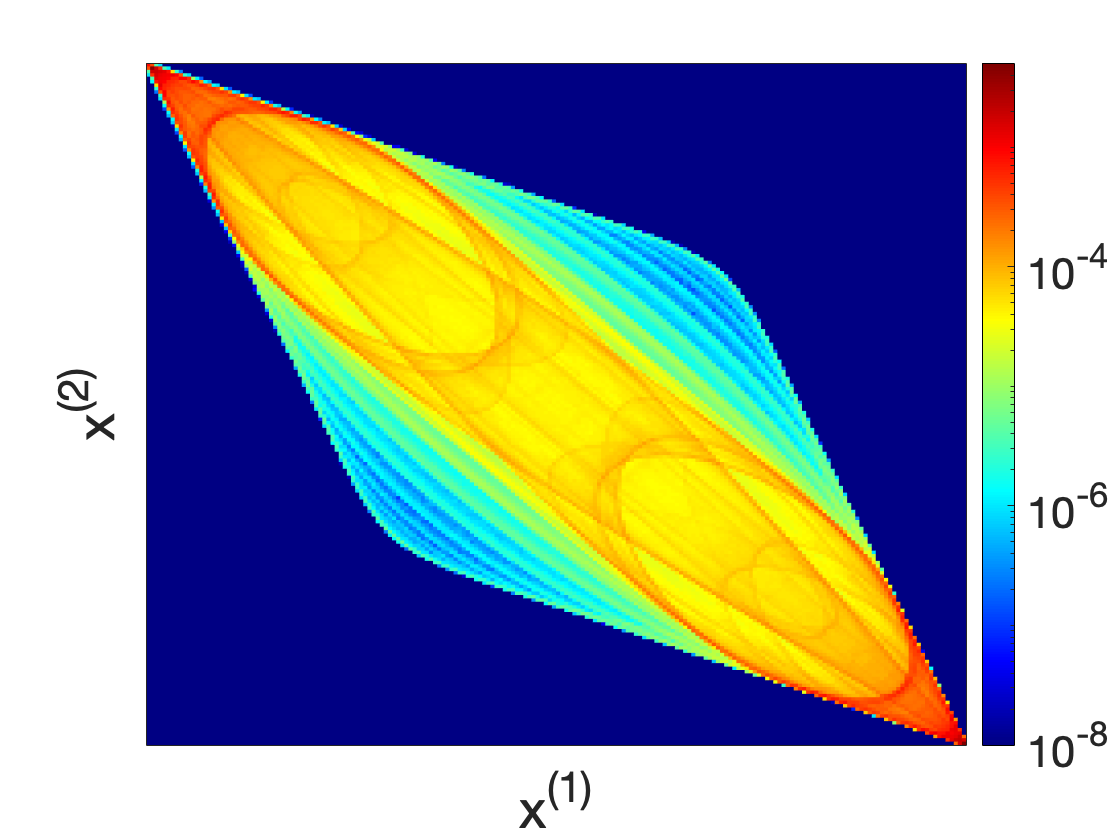} 
	\caption{$T_3$: $c = 0.29$}
\end{subfigure}
\vspace{1em}

\begin{subfigure}{0.24\linewidth}
	\includegraphics[width = \linewidth]{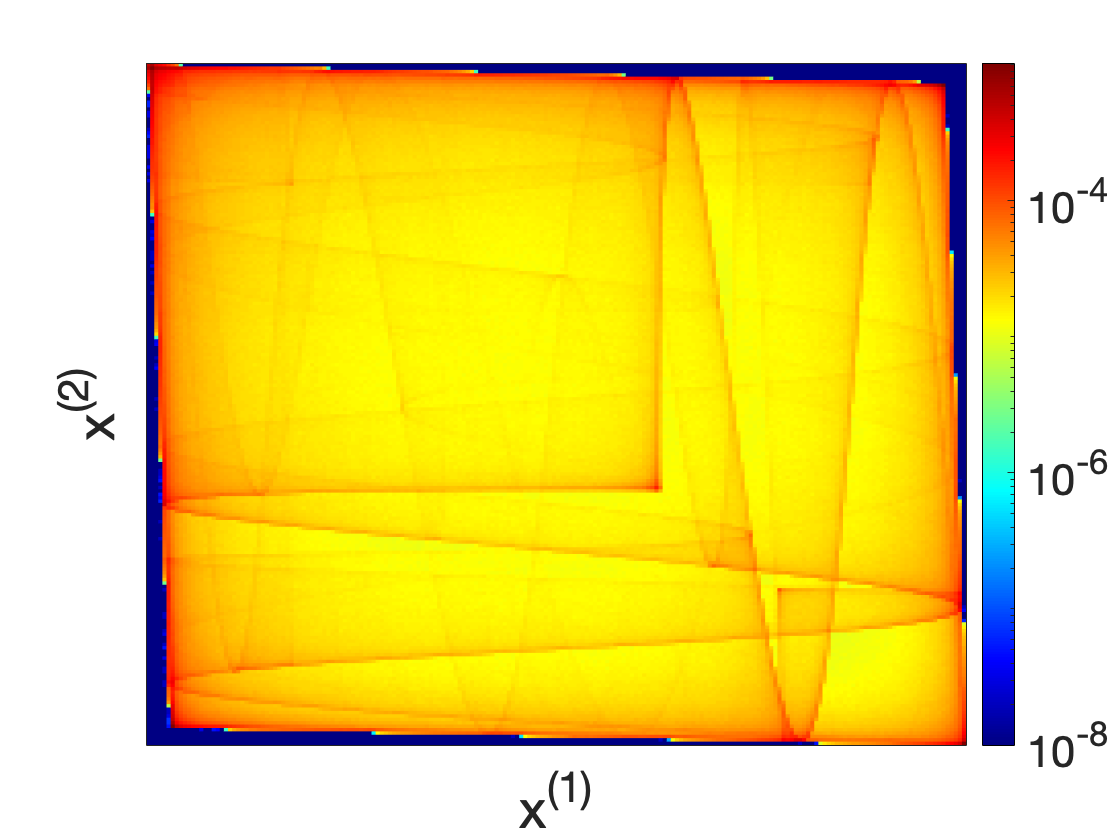}
	\caption{$T_4$: $c = 0.027$}
\end{subfigure}
\begin{subfigure}{0.24\linewidth}
	\includegraphics[width = \linewidth]{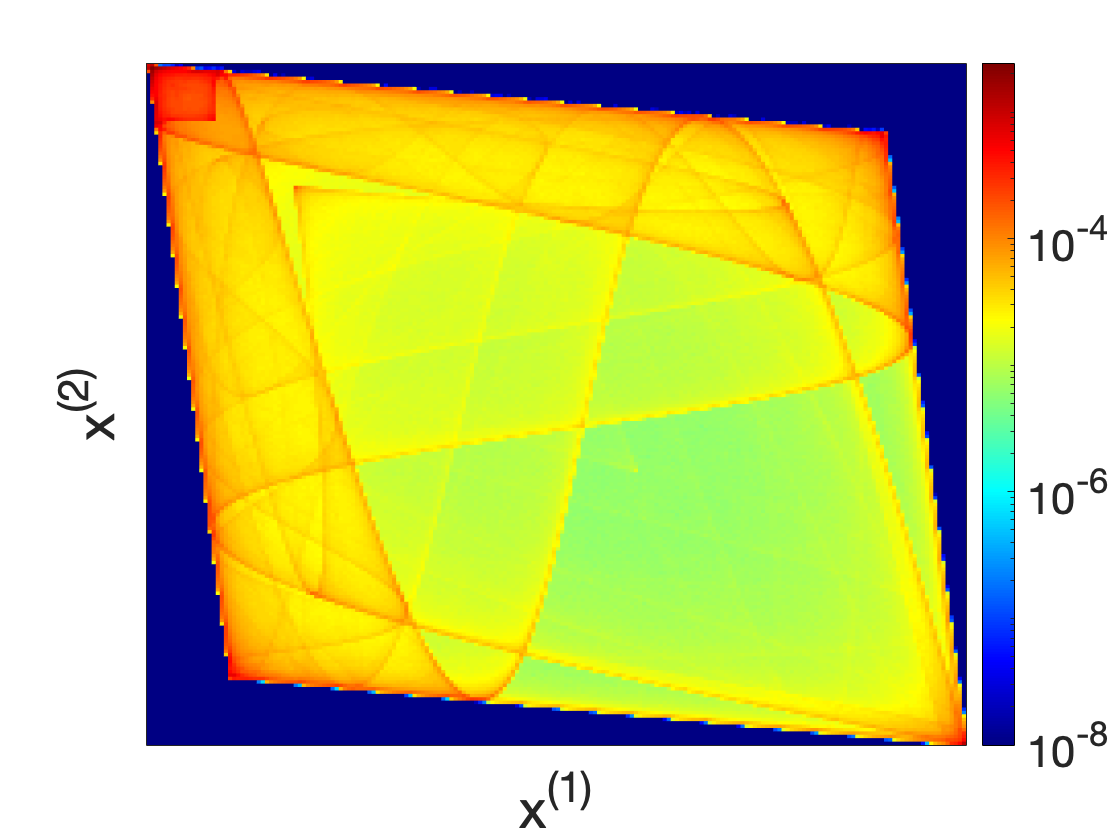} 
	\caption{$T_4$: $c = 0.1$}
\end{subfigure}
\begin{subfigure}{0.24\linewidth}
	\includegraphics[width = \linewidth]{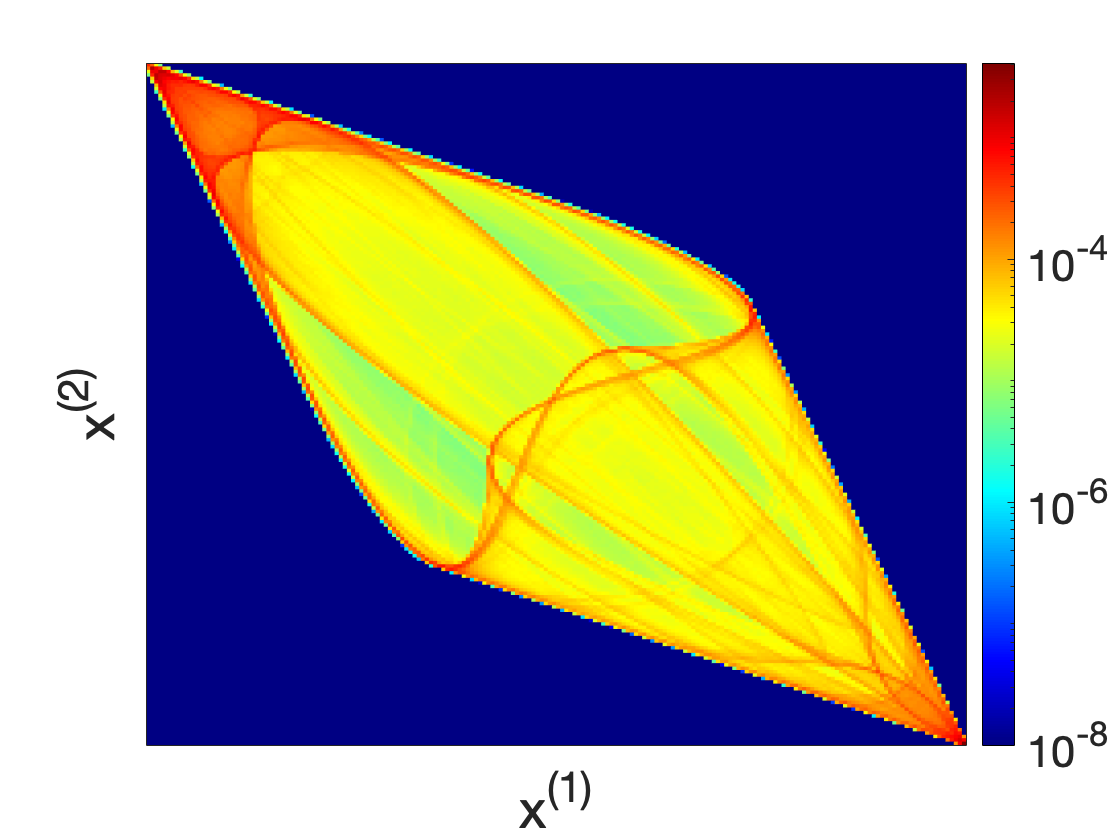} 
	\caption{$T_4$: $c = 0.29$}
\end{subfigure}
\begin{subfigure}{0.24\linewidth}
	\includegraphics[width = \linewidth]{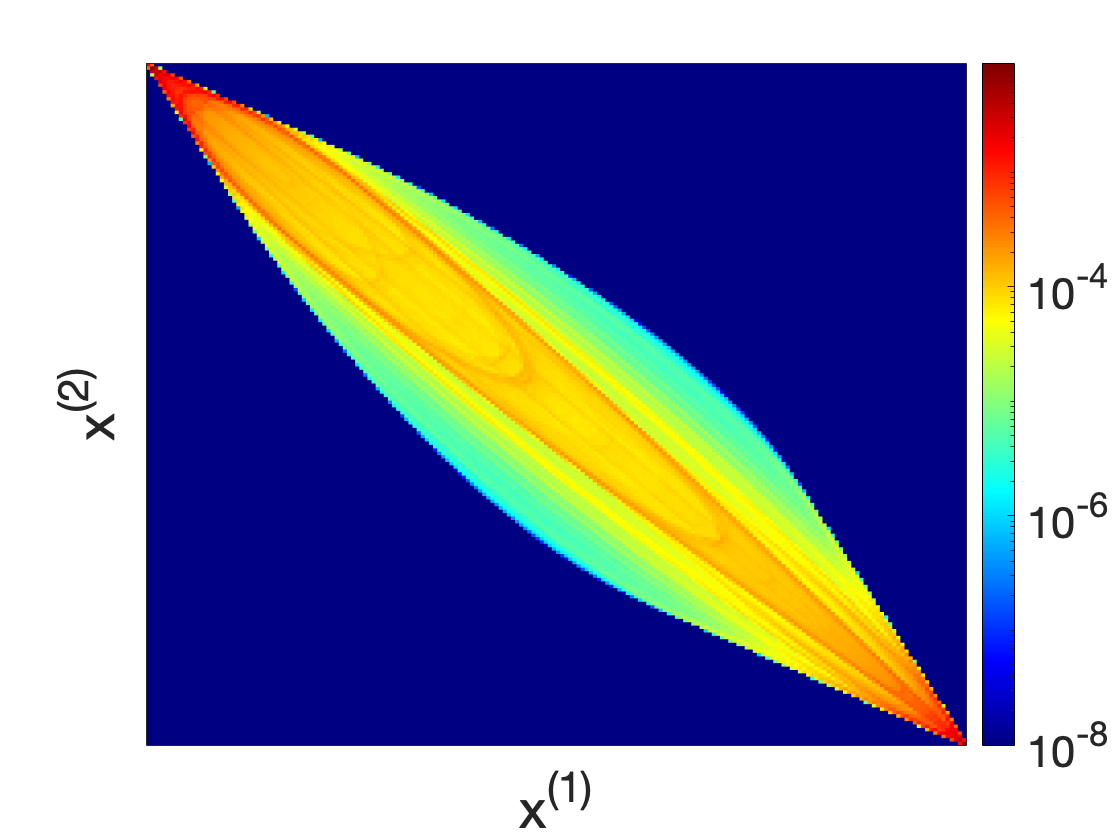} 
	\caption{$T_4$: $c = 0.35$}
\end{subfigure}
\vspace{1em}

\begin{subfigure}{0.24\linewidth}
	\includegraphics[width = \linewidth]{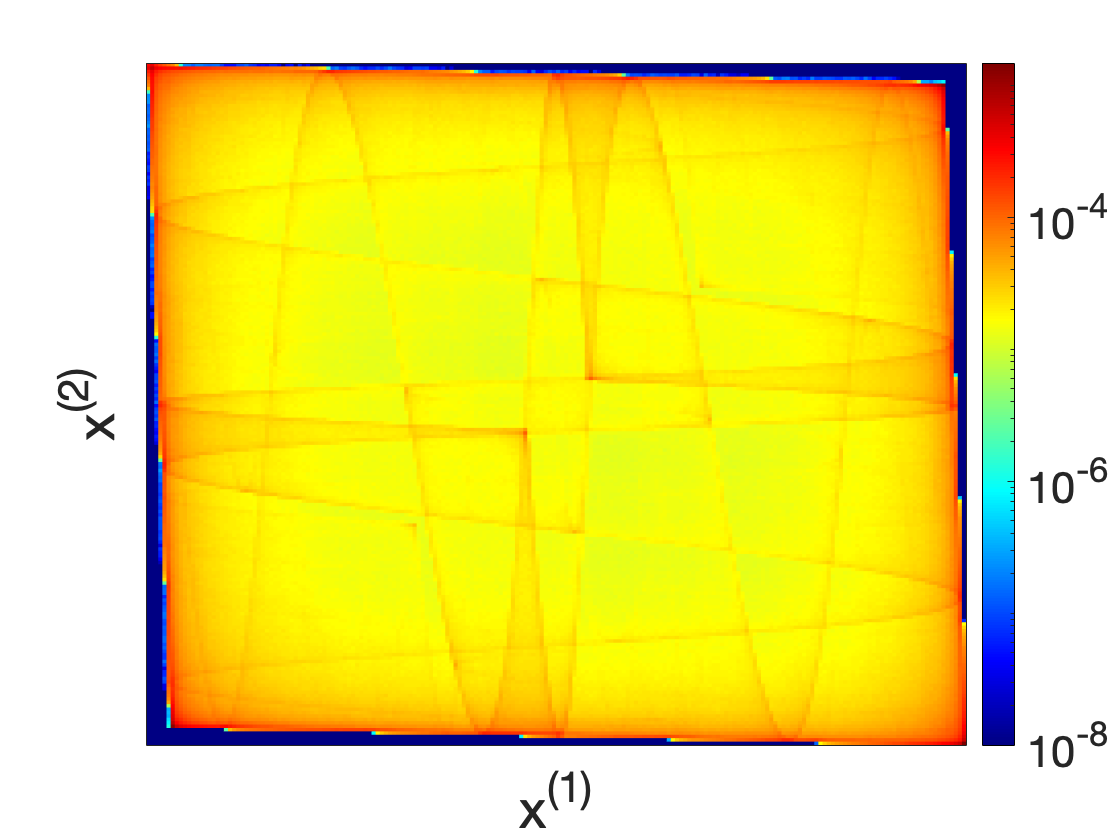}
	\caption{$T_5$: $c = 0.027$}
\end{subfigure}
\begin{subfigure}{0.24\linewidth}
	\includegraphics[width = \linewidth]{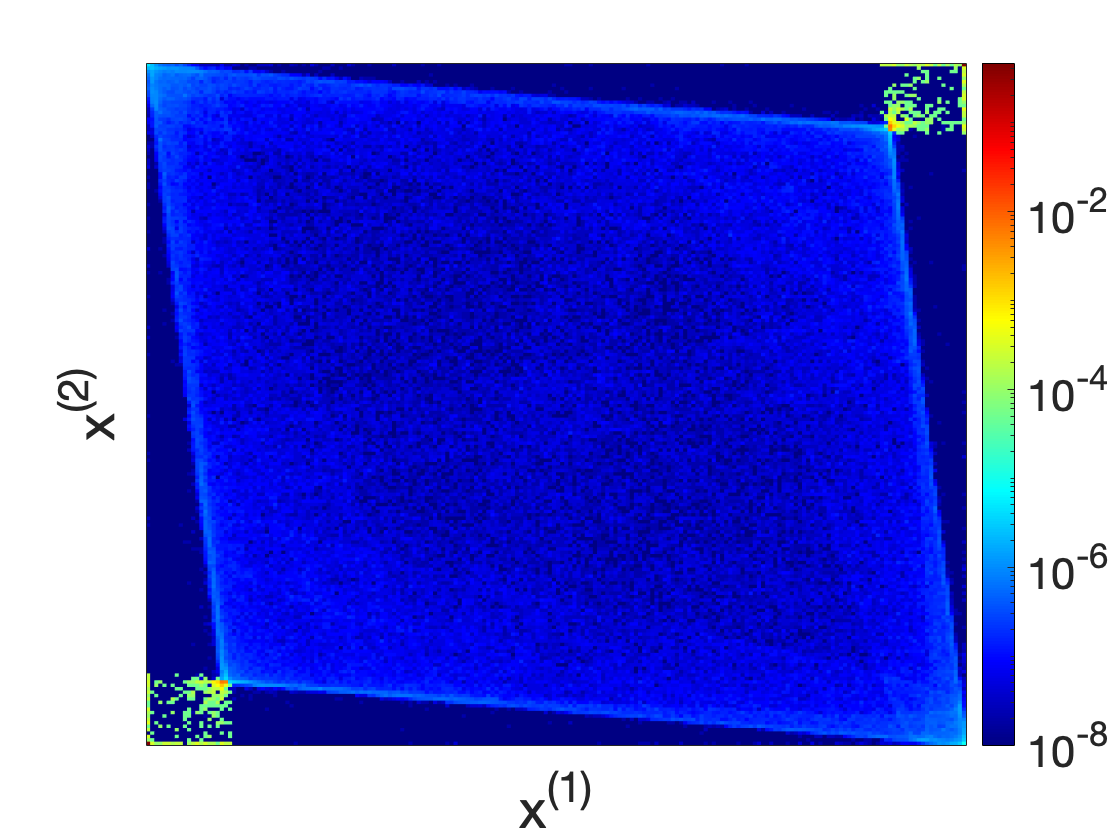} 
	\caption{$T_5$: $c = 0.093$}
\end{subfigure}
\begin{subfigure}{0.24\linewidth}
	\includegraphics[width = \linewidth]{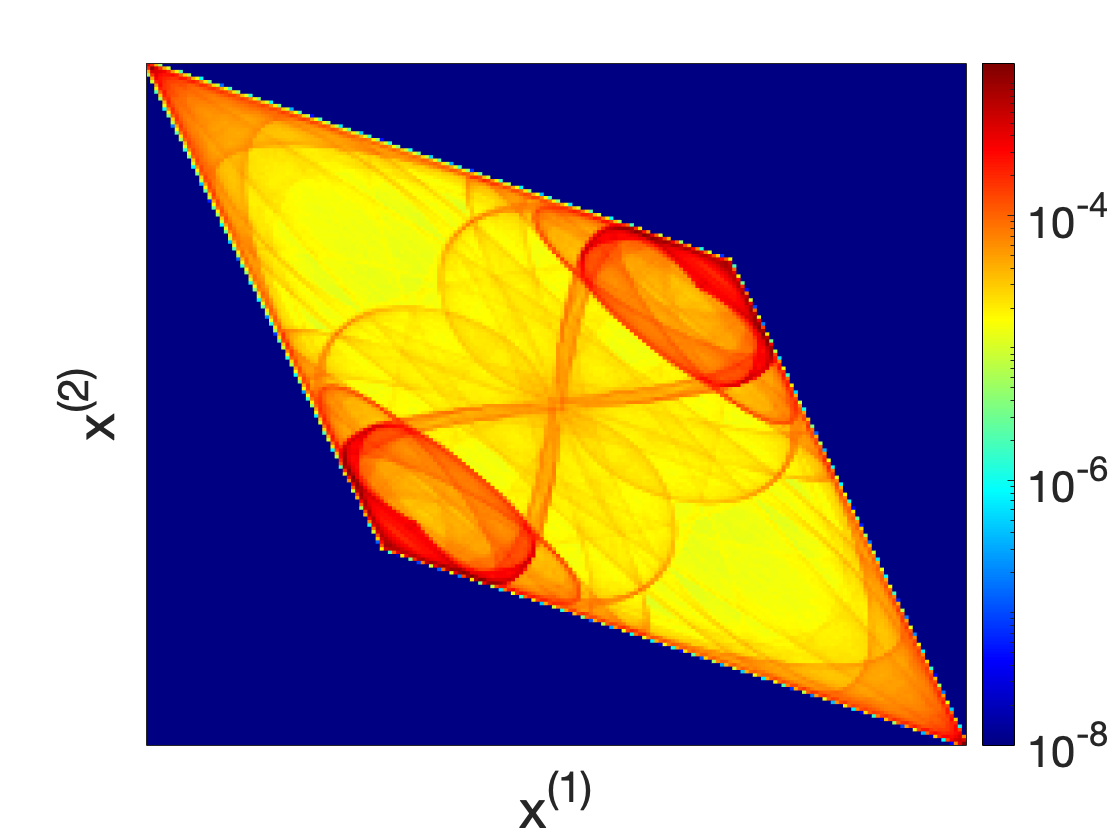} 
	\caption{$T_5$: $c = 0.29$}
\end{subfigure}
\begin{subfigure}{0.24\linewidth}
	\includegraphics[width = \linewidth]{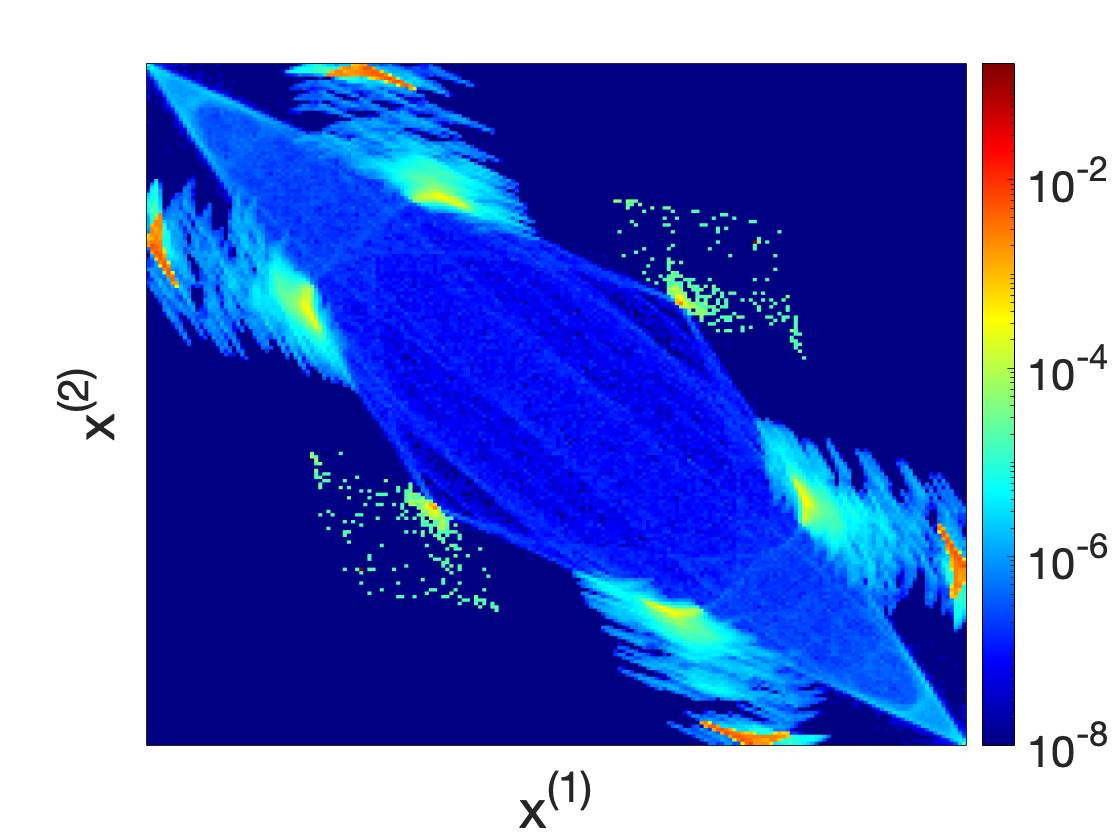} 
	\caption{$T_5$: $c = 0.35$}
\end{subfigure}
\caption{Probability densities on attractors of two coupled $T_2$ (row 1), $T_3$ (row 2), $T_4$ (row 3) and $T_5$ (row 4) systems, with different coupling strengths $c$ (indicated in each caption). The $x$- and $y$-axes intervals are $x^{(1)} \in [-1, 1]$ (value increases horizontally from left to right) and $x^{(2)} \in [-1, 1]$ (value increases vertically from top to bottom). Colour codes the averaged probability of 10,000 trajectories, randomly chosen initially, each iterated for 10,100($-100$ transient) time steps. The colour coding is in log scaling for a better representation; dark blue indicates lower probability and dark red indicates higher probability; each heatmap uses $200\times 200$ bins. Videos of the shown dynamics are available \href{https://sites.google.com/view/jinyan-cmls/home}{online.}} 
\label{2cml}
\end{figure}

For a system of two ($J = 2$) coupled $T_N$ maps, Fig.\ref{2cml} shows the attractors and the associated probability measures (in colour) for $N = 2, 3, 4$ and $5$ with various values of the coupling strength $c$. 
In general, the attractor shrinks to the diagonal ($x^{(1)} = x^{(2)}$) as the coupling strength $c$ increases, where the system achieves a completely synchronized state, before which spikes (small regions in red) occur at various $c$ values indicating high concentrations of probability; a pair of symmetric islands implies that the two lattice sites have opposite-sign dynamics ($x^{(1)}x^{(2)} < 0$). 
\\
Apart from their intricate and complicated structures, one notices that for odd $N$, the attractor is symmetric also with respect to the anti-diagonal $x^{(1)} = -x^{(2)}$ while for even $N$ this symmetry is not present. 
This is simply due to the fact that $T_N(x)$ is an odd function when $N$ is odd: $T_N(-x) = -T_N(x)$, the coupled map equation \eqref{cml-eq} still holds if we multiply the variables by $-1$, while this is not true for even $N$. 
The videos for Fig.\ref{2cml}, showing the dynamics of the attractor when the coupling strength $c$ is slowly changed, are available online \cite{videos}.

\begin{figure}[H]
\captionsetup[subfigure]{justification=centering}
\centering 
\begin{subfigure}{0.24\linewidth}
	\includegraphics[width = \linewidth]{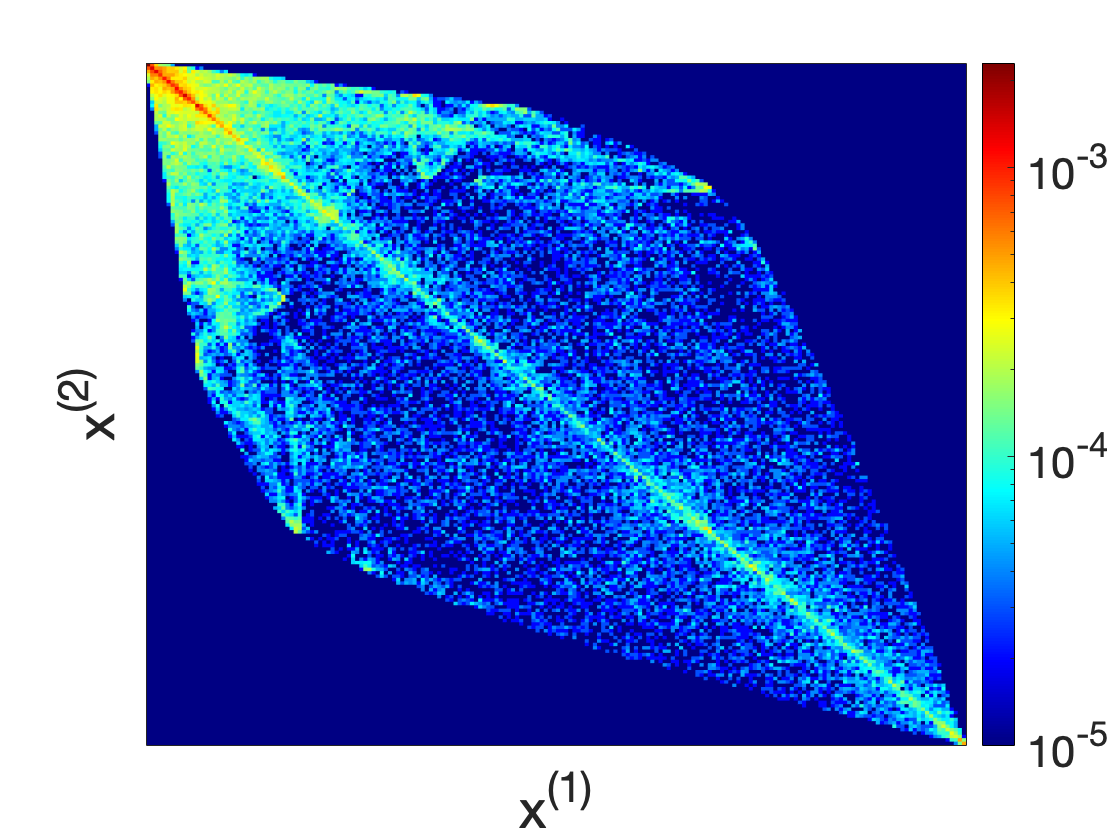}
	\caption{$T_2$: $c = 0.22$}
\end{subfigure}
\begin{subfigure}{0.24\linewidth}
	\includegraphics[width = \linewidth]{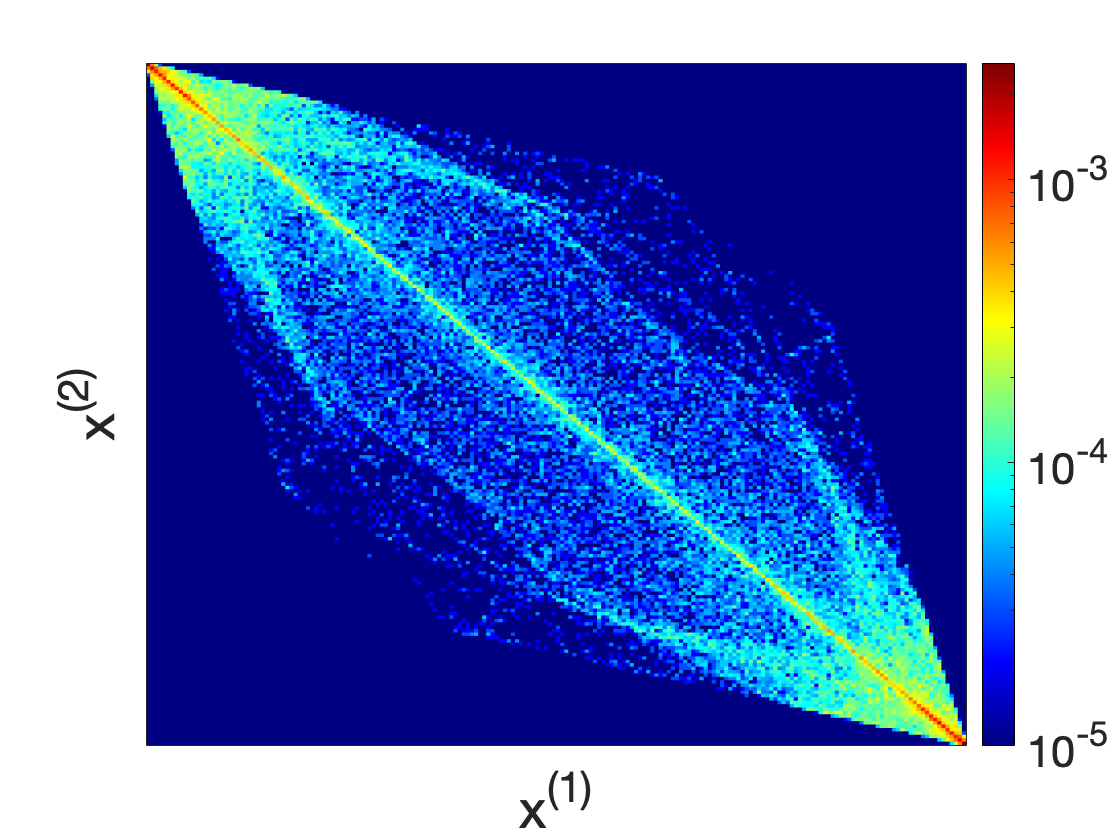}
	\caption{$T_3$: $c = 0.35$}
\end{subfigure}
\begin{subfigure}{0.24\linewidth}
	\includegraphics[width = \linewidth]{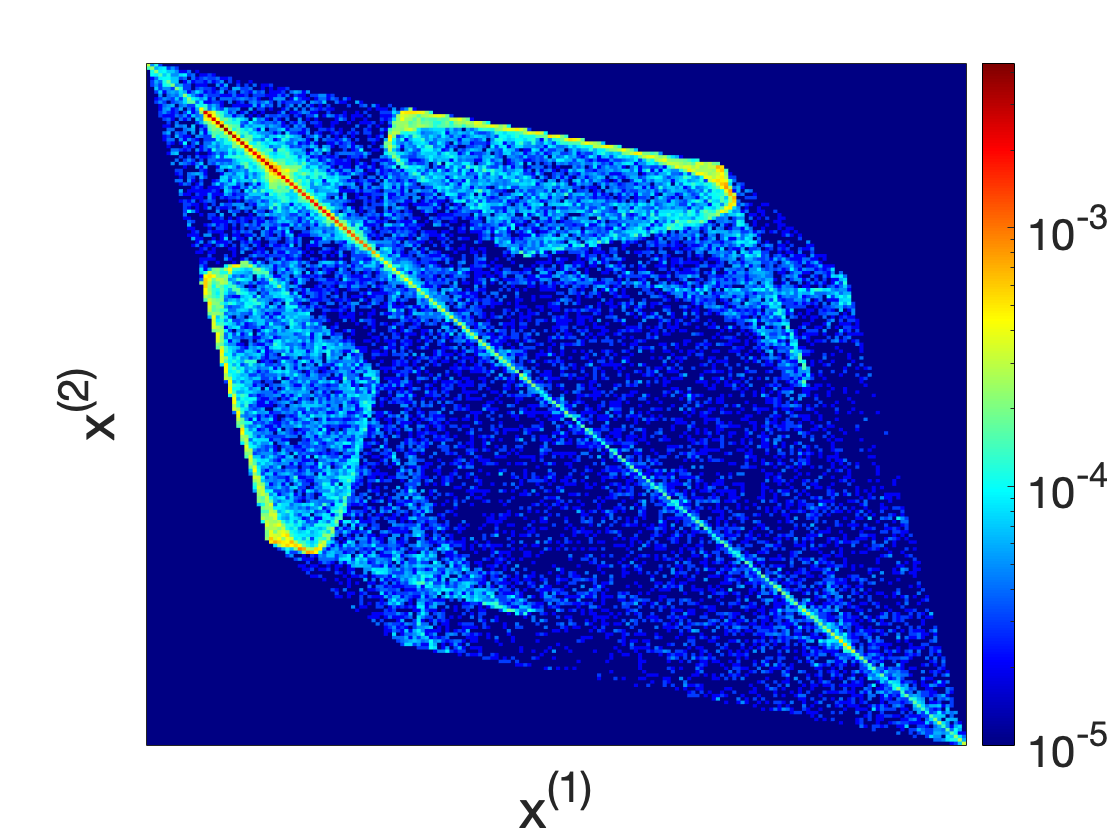}
	\caption{$T_4$: $c = 0.3$}
\end{subfigure}
\begin{subfigure}{0.24\linewidth}
	\includegraphics[width = \linewidth]{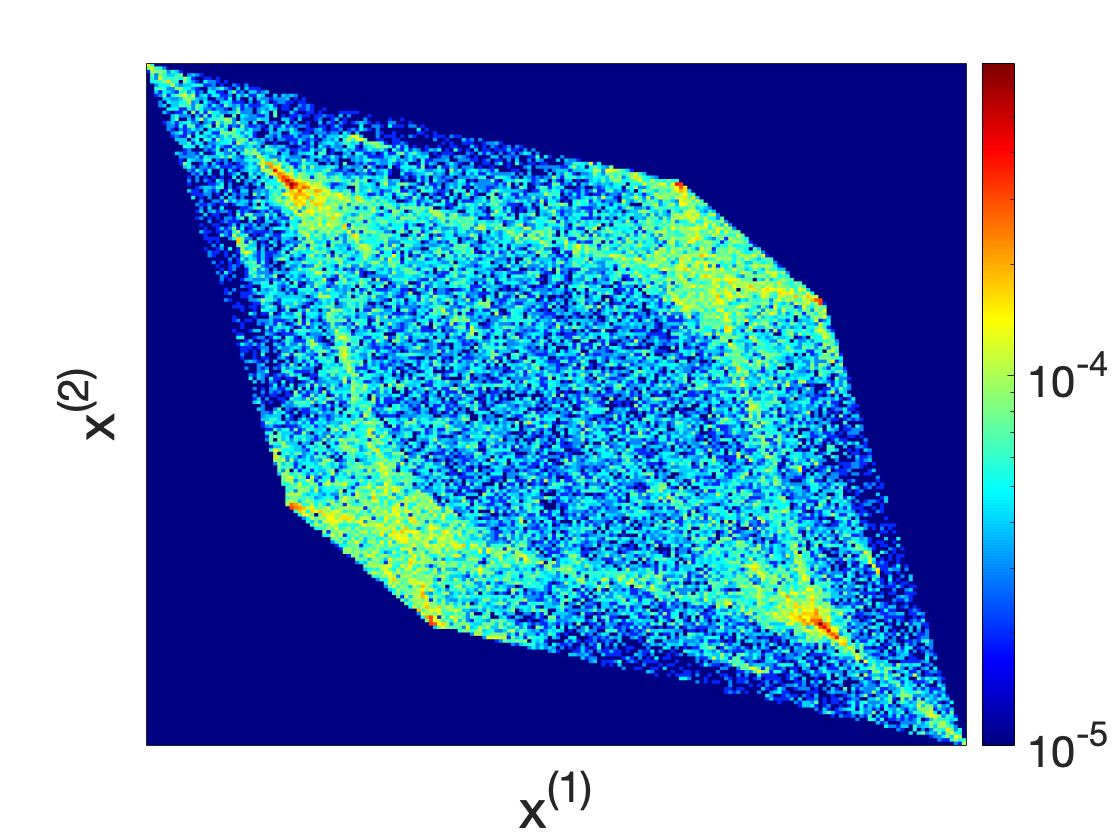}
	\caption{$T_5$: $c = 0.35$}
\end{subfigure}
\caption{Probability densities on attractors of three coupled $T_N$ systems projected onto the $(x^{(1)}, x^{(2)})$-plane of the full $(x^{(1)}, x^{(2)}, x^{(3)})$-space (i.e, marginal densities are shown). The value of the coupling strength $c$ is indicated in each caption. The axes and color coding are the same as in Fig.\ref{2cml}. The averaged probability is obtained from $500$ trajectories, randomly chosen initially, each iterated for $300$($-100$ transient) time steps.}
\label{3cml}
\end{figure}

Fig.\ref{3cml} shows similar symmetry behaviour for a system of three ($J = 3$) coupled $T_N$ maps of order $N = 2, 3, 4$ and $5$ with some arbitrarily chosen coupling strength $c$ as indicated in each caption. Each three-dimensional attractor is projected onto the $(x^{(1)}, x^{(2)})$-plane, in other words, the marginal density $\int_{-1}^1 \rho(x^{(1)}, x^{(2)}, x^{(3)}) dx^{(3)}$ is shown. 

In \cite{book, physica-d} four types of coupled map systems were introduced for applications in stochastically quantized field theories. They differ from the current model (eq.\eqref{cml-eq}, so-called {\it forward diffusive coupling}) by changing the sign of the second term ({\it antidiffusive}) or replacing $T_N(x^{(j \pm 1)}_n)$ by $x^{(j \pm 1)}_n$ ({\it backward coupling}, taking into account the propagation time along the lattice). All of them have interesting physical applications but in this paper we focus on the model eq.\eqref{cml-eq} only. 
In the next section, we show that in this simple model of coupled information shifts there might be a deep connection to a fundamental constant of particle physics, the fine structure constant. For more details on this type of approach fixing standard model parameters, see \cite{book, physica-d}.

\section{Vanishing spatial correlations related to the fine-structure constant}
\label{sect-snnc}
For applications in quantum field theory and high energy physics it is meaningful to extend the coupled map system to a larger size, and ultimately to take the thermodynamic limit $J \to \infty$. 

In this section we study numerically the {\it spatial} nearest-neighbour correlations (SNNC) of the system described by eq.\eqref{cml-eq}, and show that the value of the fine structure constant is distinguished as a coupling constant for which the SNNC (with local $T_3$ maps) vanishes, implying that the underlying dynamics is uncorrelated in the spatial direction, thus achieving maximal randomness. 

The spatial nearest-neighbour correlation (SNNC) for the coupled map system \eqref{cml-eq} of size $J$ (large) is defined as 
\begin{equation*}
\text{SNNC} := \lim_{K \to \infty} \frac{1}{KJ} \sum_{n = 1}^K \sum_{j = 1}^J x_n^{(j)}x_n^{(j+1)}, 
\end{equation*}
where the superscripts $j$ and subscripts $n$ of $x$ denote the lattice location and iteration times, respectively as before.

\begin{figure}[H]
\captionsetup[subfigure]{justification=centering}
\centering 
\begin{subfigure}{0.48\linewidth}
	\includegraphics[width = \linewidth]{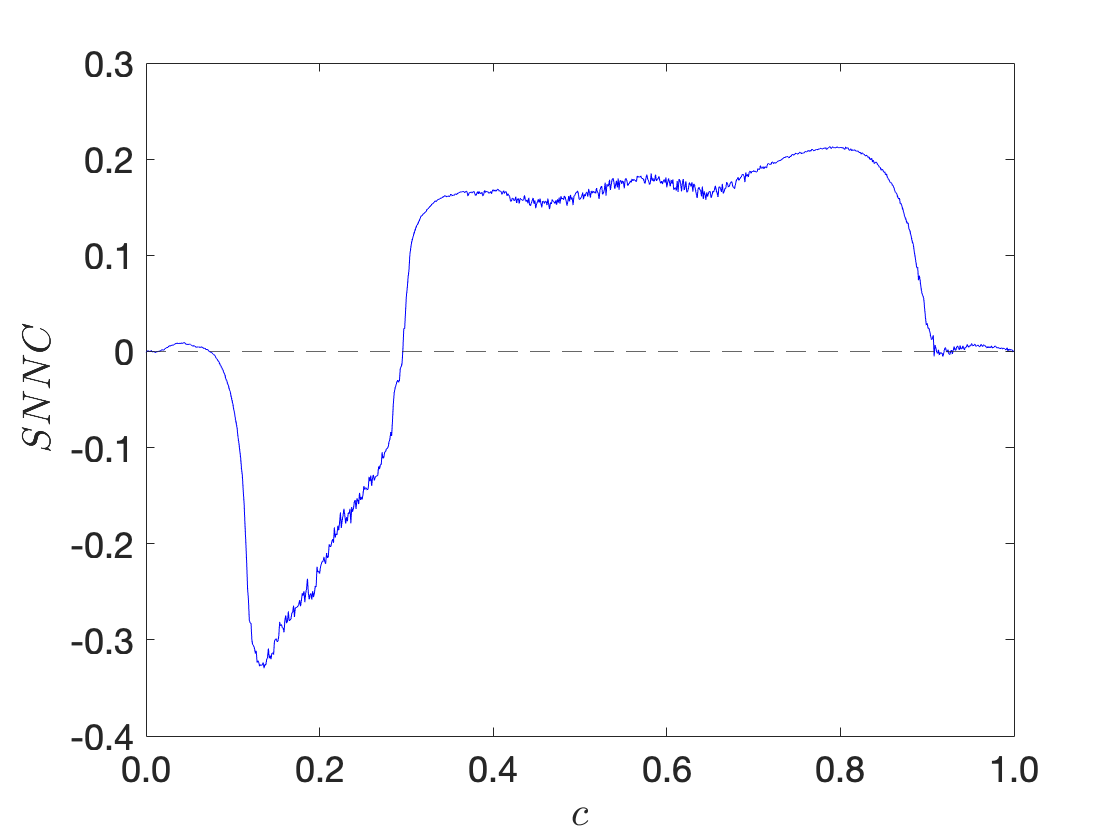}
	\caption{SNNC for $c \in [0, 1)$}\label{snnc-T3-c0to1}
\end{subfigure}
\begin{subfigure}{0.48\linewidth}
	\includegraphics[width = \linewidth]{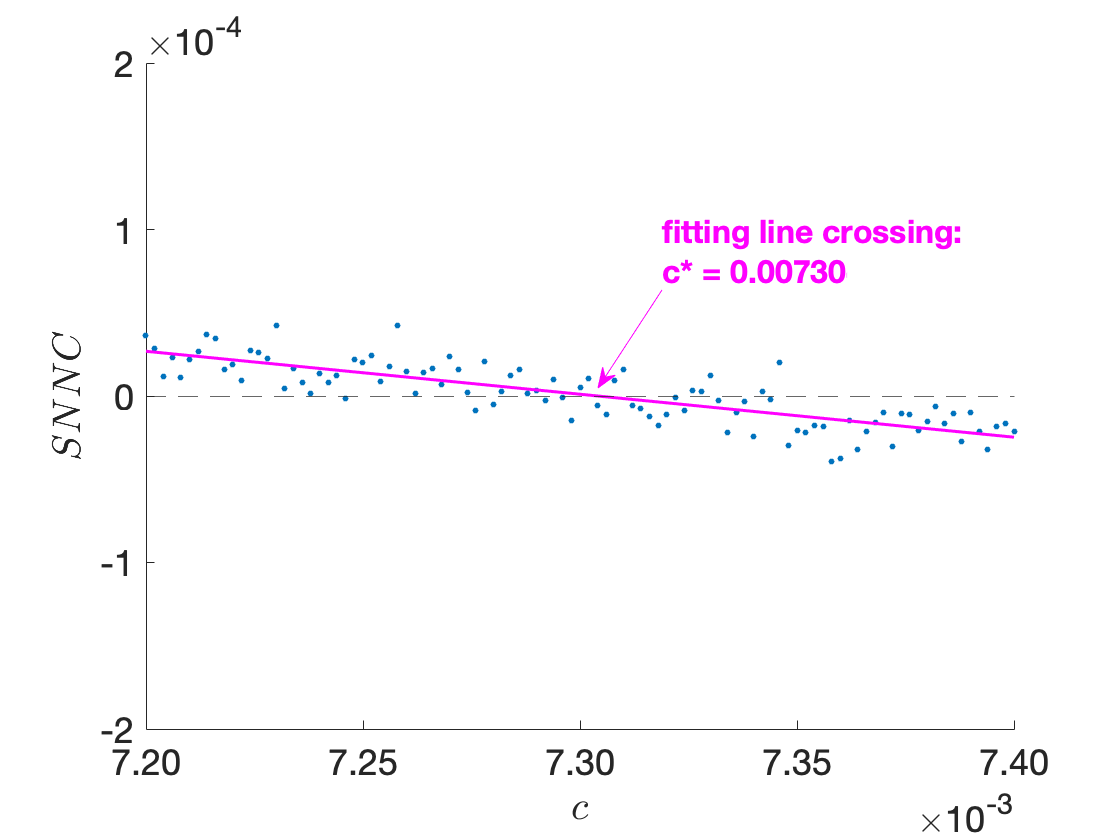}
	\caption{zoomed-in for $c \in [0.0072, 0.0074]$}\label{snnc-T3-zoomed}
\end{subfigure}
\caption{The graph of spatial nearest-neighbour correlation (SNNC) for coupled 3rd-order Chebyshev maps ($T_3$) as a function of the coupling strength $c \in [0, 1)$. The space$\times$time size for the first plot is $1000\times 1100$($-100$ for transient), and for the second plot is $50000\times 50100$($-100$); all initial points are randomly chosen from a uniform distribution in the interval $[-1, 1]$.}
\label{snnc-T3}
\end{figure}

Here we choose the local Chebyshev map $T_N$ of order $N = 3$, the reader can refer to \cite{jycb20, book} for results on other orders. Fig.\ref{snnc-T3} shows our results for SNNC as a function of the coupling $c$, as obtained from new numerical simulations. One notices in Fig.\ref{snnc-T3-c0to1} that for the coupling strength $c$ very small there is a zero of SNNC around $c \in [0.0072, 0.0074]$, and by magnifying this region in Fig.\ref{snnc-T3-zoomed} with higher precision, we observe that the zero is located at a particular value $c^*$. Our simulations were repeated 11 times for different random initial conditions and the averaged result is $c^* = 0.0073084(48)$, where the statistical error is estimated from the small deviations that are observed for each of the 11 runs. This value is in good agreement with the numerical value of the fine structure constant $\alpha_{el} \approx 1/137 =0.00730$ as observed in nature. In fact, the fine structure constant $\alpha_{el}(E)$ is a running coupling constant and depends on the energy scale $E$ considered \cite{book}. Our numerical results are consistent with the range $m_e \leq E \leq 3m_e$, where $m_e$ is the electron mass. Our simulations were performed on lattices of size $J=50000$ with periodic boundary conditions. The iteration time for each run was $K=50100$, with 100 initial iterations discarded to avoid transients. At smaller lattice sizes $J$, we observe a small tendency of the zero $c^*$ to decrease with increasing lattice size, so there may also be tiny systematic errors due to the finite lattice size used in our simulations.

The general idea of this approach is that in a pre-universe, before the creation of 4-dimensional space-time,
the coupled information shift dynamics is already evolving and fixing standard model parameters such as the fine structure constant as states of maximum randomness (realized as a state of zero SNNC), given the constraint that we have a deterministic (chaotic) dynamics on microscopic scales.

More extensive numerical simulations can in principle be performed on supercomputers to locate the exact numerical value of $c^*$, making the result more precise. Within the precision obtained in our simulations, the coupled Chebyshev system (with local $T_3$ maps) gives a possible distinction of a fundamental constant of particle physics, by making the coupled chaotic system uncorrelated, for which the deterministic system appears completely random (in the spatial direction) in the sense of vanishing nearest-neighbour correlations. 

Of course, one might argue that perhaps the value of the zero-crossing in Fig.~4b is just a random coincidence, i.e. just by chance happens to coincide with the observed value of the fine structure constant. However, coincidences with other values of standard model coupling constants (weak and strong interactions) were also observed for other types of chaotic strings, see \cite{book,physica-d,prd} for more details. Thus a pure random coincidence of all these observed values appears quite unlikely, see \cite{book} for quantitative estimates of the likelihood. 

One can also study in a similar way the {\it temporal} nearest-neighbour correlations (TNNC), see \cite{jycb20}, and also other higher-order statistics. Our results seem to indicate that the distinction comes only from the spatial two-point correlation function,
not from other types of higher-order correlation functions.

\section{Conclusion}
In this paper we have considered an information shift dynamics which is deterministic and chaotic, and which maximizes Tsallis $q=3$ entropy in a generalized statistical mechanics formalism. It is, in a sense, the simplest example of a low-dimensional statistical mechanical system that lives on a compact phase space and is ergodic and mixing.
We consider it as a candidate dynamics for a pre-universe information shift dynamics that has the ability to fix standard model parameters (such as the fine structure constant) if a coupled version is considered.


\end{document}